\documentclass[namedreferences]{solarphysics}

\usepackage[hyperref,optionalrh]{spr-sola-addons} 
\usepackage[optionalrh]{spr-sola-addons} 
\usepackage{graphicx}        
\usepackage{amssymb}        
\usepackage{color}  
\usepackage{pdflscape}	
\usepackage{breakurl}        

\chardef\us=`\_

\begin{document}

\begin{article}
\begin{opening}

\title{On modeling ICME cross-sections as static MHD columns}

\author[addressref={aff1}, corref, email={debesh.bhattacharjee@students.iiserpune.ac.in}]{\inits{D.B.}\fnm{Debesh}~\lnm{Bhattacharjee}}
\author[addressref=aff1,email={p.subramanian@iiserpune.ac.in}]{\inits{P.S.}\fnm{Prasad}~\lnm{Subramanian}}
\author[addressref=aff2]{\inits{V.B}\fnm{Volker}~\lnm{Bothmer}}
\author[addressref=aff3]{\inits{T.NC}\fnm{Teresa}~\lnm{Nieves-Chinchilla}}
\author[addressref=aff4]{\inits{A.V}\fnm{Angelos}~\lnm{Vourlidas}}

\address[id=aff1]{Indian Institute of Science Education and Research, Pune ;
Dr. Homi Bhabha Road, Pashan, Pune 411008,
India}
\address[id=aff2]{Institut f{\"u}r Astrophyzik , Georg-August Universit{\"a}t G{\"o}ttingen, G{\"o}ttingen, Germany}
\address[id=aff3]{Heliophysics science division, NASA-Goddard Space Flight Center, Greenbelt, MD (USA).}
\address[id=aff4]{The Johns Hopkins University Applied Physics Laboratory, Laurel MD, USA.}

\runningauthor{Bhattacharjee et al.}
\runningtitle{Magnetohydrostatic assumption for ICMEs}

\begin{abstract}
Solar coronal mass ejections are well known to expand as they propagate through the heliosphere. Despite this, their cross-sections are usually modeled as static plasma columns within the magnetohydrodynamics (MHD) framework. We test the validity of this approach using {\em in-situ} plasma data from 151 magnetic clouds (MCs) observed by the WIND spacecraft and 45 observed by the Helios spacecrafts. We find that the most probable cross-section expansion speeds for the WIND events are only $\approx 0.06$ times the Alfv\'en speed inside the MCs while the most probable cross-section expansion speeds for the Helios events is $\approx 0.03$. MC cross-sections can thus be considered to be nearly static over an Alfv\'en crossing timescale. Using estimates of electrical conductivity arising from Coulomb collisions, we find that the Lundquist number inside MCs is high ($\approx 10^{13}$), suggesting that the MHD description is well justified. The Joule heating rates using our conductivity estimates are several orders of magnitude lower than the requirement for plasma heating inside MCs near the Earth. While the (low) heating rates we compute are consistent with the MHD description, the discrepancy with the heating requirement points to possible departures from MHD and the need for a better understanding of plasma heating in MCs.
\end{abstract}
\end{opening}

\section{Introduction}
\label{S-Introduction}

Coronal Mass Ejections (CMEs) are magnetized plasma structures
erupting from the solar corona \citep{2011ChenLRSP}. Some of these CMEs have been sampled {\em in-situ} by several spacecraft, yielding detailed information on the CME plasma parameters along the line of intercept of the spacecraft. This has resulted in an extensive database of observations of the interplanetary counterparts of CMEs, which are called ICMEs \citep{2007ZhangJGRA}. The cross-sectional structure of ICMEs is typically modeled as a static plasma column, using the framework of magnetohydrodynamics (MHD) - this includes early works such as \citet{51LqPhRv} , \citet{88BJGR} to works that are based on Grad-Shafranov equation \citep{2002HuJGRA, 2009MoSoPh, 2011IvSoPh} and similar formalisms \citep{2016NCApJ, 2018NCApJ, 2002CidSoPh}. This is similiar in philosophy to analyze of the equilibrium of plasma columns in laboratory settings; e.g., \citet{2003Boyd.Sanderson}.

It is well known that ICMEs expand as they travel through the heliosphere, suggesting that they are dynamic, {\bf expanding} structures; e.g., \citet{1998BthAnGeo,2009VASTRA,2010GulisanoA&A}.  However, as mentioned above, ICME cross-sections seem to be quite well described as static plasma columns. It is therefore worth examining how the apparent success of such static models can be reconciled with the observation of expanding ICME cross-sections. This paper concentrates on the following broad areas:

\begin{enumerate}

\item
Although ICME cross-sections are observed to expand, how do the expansion speeds compare with typical signal propagation speeds (such as the Alfv\'en speed) inside the plasma? If the expansion speeds are slow, the concept of a static column would be justified.

\item

We check if the electrical conductivity of the medium is large (it is technically infinite in MHD), which means that the magnetic diffusivity is very small. This is the basis of the well known ``frozen-in'' condition for magnetic fields in MHD. The Lundquist number, which is the ratio of the magnetic diffusion timescale to the Alfv\'en crossing timescale, is a standard measure to quantify the goodness of the frozen field assumption. The larger the Lundquist number, the better the validity of the MHD description.



\item
An infinitely conducting plasma would be non-dissipative, in disagreement with expectations of plasma heating inside ICMEs. Having computed the electrical conductivity for item 2, we use an approximate estimate of the current density to calculate the Joule heating rate and compare it with inferred MC plasma heating rates.
\end{enumerate}

The data we use is explained in \S~\ref{S - Data}. From here on, we will restrict our attention only to the magnetic clouds (MCs) within the overall ICME structure. MCs are the magnetically well structured parts of ICMEs and their boundaries and expansion speeds are typically better defined. 
We compare the observed MC expansion speeds with Alfv\'en speed in \S~\ref{S - static assumption}. We evaluate the collisional conductivity and use it to evaluate the Lundquist numbers for MCs are calculated in \S~\ref{S - Lundquist}. The conductivity is used to compute Joule heating rates in \S~\ref{S - Joule}. Our overall conclusions on the applicability of the MHD paradigm to MCs and caveats are in presented \S~\ref{S - summary}.


\section{Data}
\label{S - Data}

We use in-situ data from three different spacecraft (WIND, Helios 1 and Helios 2) for our current study. The WIND ICME catalogue (\url{https://wind.nasa.gov/ICMEindex.php}) provides a sample of well observed Earth directed ICMEs as observed by the WIND spacecraft \citep{2019NCSoPh, 18NCSo} near the earth. The magnetic clouds (MCs) associated with these ICMEs are classified into different categories depending upon how well the observed plasma parameters fit the expectations of a static flux rope configuration. Of all the ICMEs observed between the years 1995 and 2015 listed on the WIND website, we first shortlist those that are categorized as F+ and Fr events. These events best fit the expectations of the flux rope model \citep{2016NCApJ, 18NCSo}. Fr events indicate MCs with a single magnetic field rotation between $90^{\circ}$ and $180^{\circ}$ and F+ events indicate MCs with a single magnetic field rotation greater than $180^{\circ}$. We further shortlist events that are neither preceded nor followed by other ICMEs or ejecta two days before and after the event under consideration, so as to exclude interacting CMEs. Our final shortlist comprises 151 ICMEs from the WIND catalogue.

We also use in-situ data from the 45 well observed MCs from Helios observations shortlisted by \citet{1998BthAnGeo}. These 45 events, observed between December, 1974 and July, 1981, provide us an opportunity to analyze ICMEs at heliocentric distances ranging from 0.3 AU to 1 AU. This composite dataset helps us analyze the behavior of well observed MCs over a wide range of heliospheric distances. The shortlisted events from the WIND spacecraft are listed in Table \ref{S - Table A1} while those from the Helios 1 and 2 spacecrafts are listed in Table \ref{Table A2}.


\section{How good is the ``static'' assumption for ICMEs?}

\label{S - static assumption}
ICME cross-sections are modeled as {\em static} plasma columns. However, it is clearly evident from observations that they expand as they propagate. \\
Our findings show how and why the static assumption is justified. The momentum equation for ideal MHD is given by

\begin{equation}
\label{eq_momentum_conservation}
\frac{\partial\mathbf{v}}{\partial t} + \left( \mathbf{v \cdot \nabla } \right)\mathbf{v} = -\frac{1}{\rho} \nabla P + \frac{1}{4 \pi \rho} (\nabla \times \mathbf{B}) \times \mathbf{B} + \frac{\mathbf{F}}{\rho}
\end{equation}

Here $\mathbf{v}$ , $P$ , $\rho$ are the bulk plasma velocity, pressure and mass density respectively. The second term on the right hand side (RHS) of Eq~\ref{eq_momentum_conservation} represents the Lorentz self-force $(1/c) \, {\mathbf j} \times {\mathbf B}$, where ${\mathbf j} \equiv (c/4 \pi) (\nabla \times \mathbf{B})$ is the current density. The last term on the RHS involving $\mathbf{F}$ represents external forces such as gravity. 
If the system is in magnetohydrostatic equilibrium, the plasma velocity $\mathbf{v} = 0$ [e.g., \S~14.3, \citet{1998choudhuriAR_physics}; \S~4.3 \citet{2003Boyd.Sanderson}]. The gravitational attraction due to the Sun becomes negligible beyond a few solar radii above the photosphere, and ${\mathbf F}$ can therefore be ignored (\S~2, \citet{2004CargillSoPh}). Eq~\ref{eq_momentum_conservation} for magnetohydrostatic equilibrium therefore reduces to \citep{2016NCApJ}
\begin{equation}
\label{eq_static}
\frac{1}{4 \pi \rho} (\nabla \times \mathbf{B}) \times \mathbf{B} = \frac{1}{\rho} \nabla P
\end{equation}
We note that Eq~\ref{eq_static} can be obtained from Eq~\ref{eq_momentum_conservation} if the plasma velocity ${\mathbf v} = 0$ and/or if the material derivative of the plasma velocity $D{\mathbf v}/Dt \equiv \partial {\mathbf v}/\partial t + ({\mathbf v} \, . \, {\mathbf \nabla}) {\mathbf v} = 0$. In our context, magnetohydrostatic equilibrium refers to the first assumption (${\mathbf v} = 0$). Equation \ref{eq_static} depicts the balance between the Lorentz force and the force due to the pressure gradient in the plasma. Early models \citep{88BJGR, 1990LeppingJGR} assumed that the flux rope would be force-free (i.e., ${\mathbf j} \times {\mathbf B} = 0$), in which case the RHS of Eq~\ref{eq_static} would be zero. Some later models; e.g., \citet{2009MoSoPh, 2019ScoA&A}.  \citet{2016NCApJ} relax this assumption and use all of Eq~\ref{eq_static} (i.e., the RHS involving ${\mathbf \nabla} P$ is not assumed to be zero). 

\begin{figure}    
   \centerline{\hspace*{0.015\textwidth}
               \includegraphics[width=0.515\textwidth , scale=0.6,width=7.5cm, height=5.0cm]{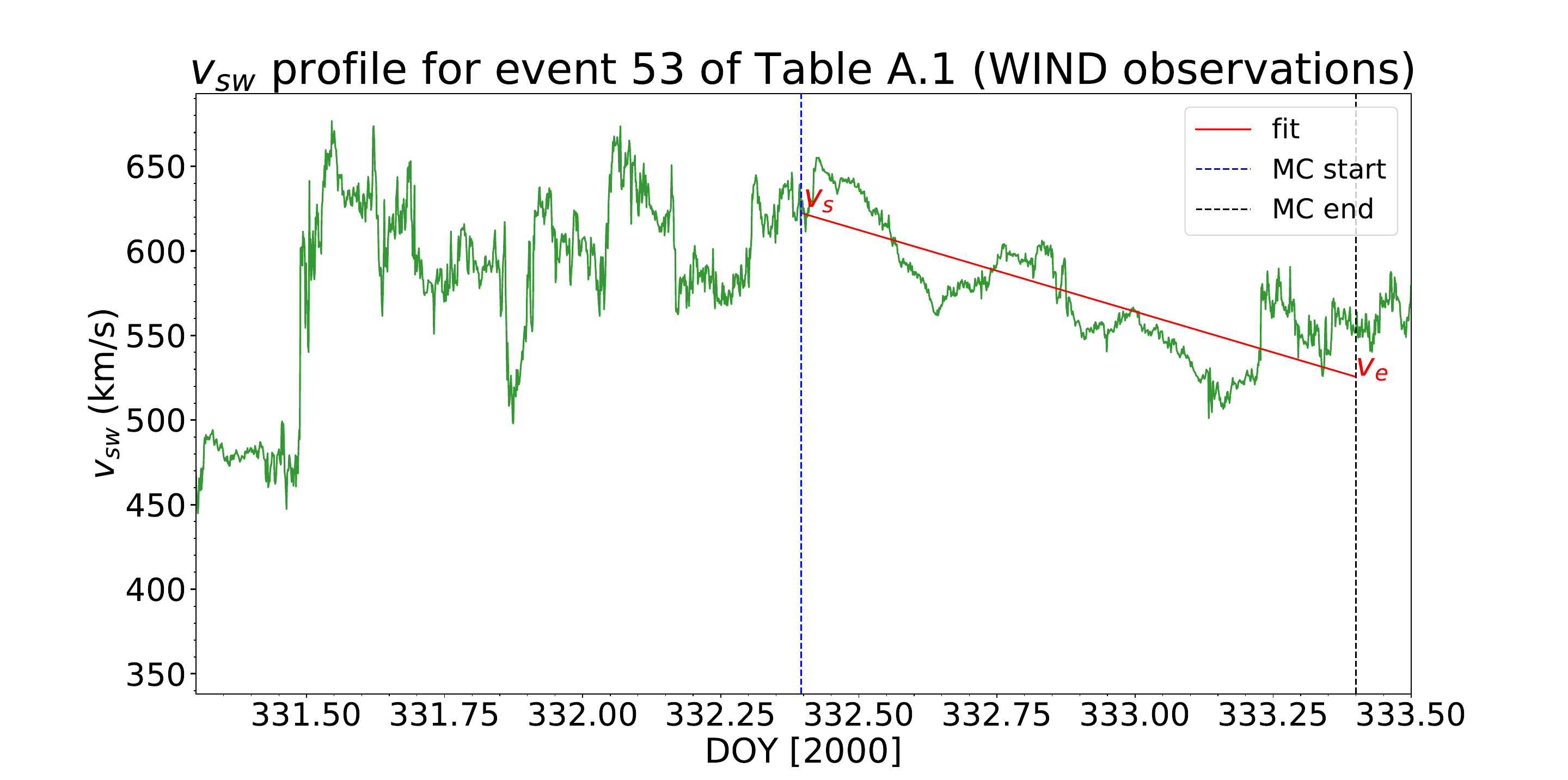}
               \hspace*{-0.03\textwidth}
               \includegraphics[width=0.515\textwidth , scale=0.6,width=7.5cm, height=5.0cm]{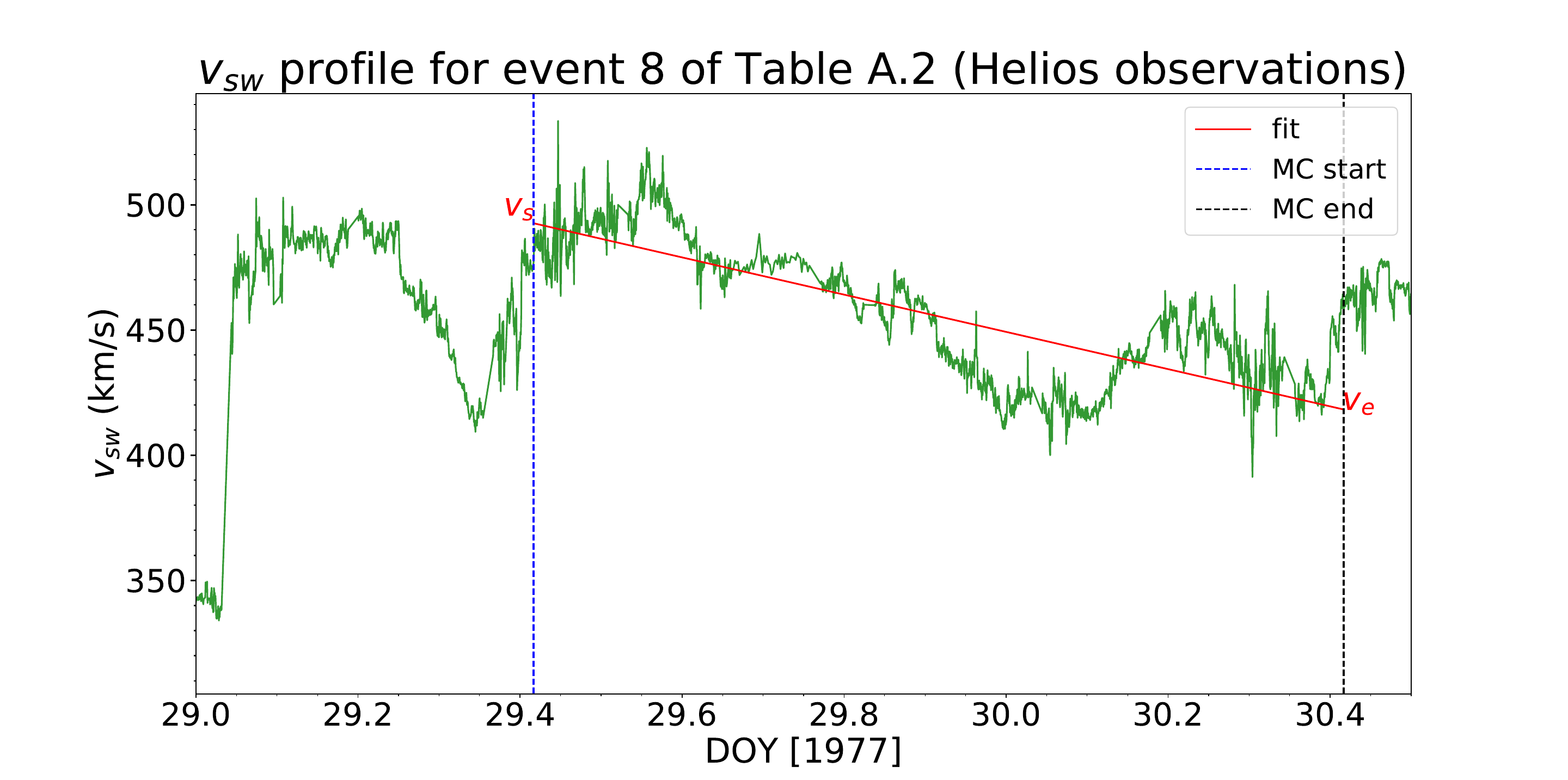}
              }
     \vspace{-0.35\textwidth}   
     \centerline{\Large \bf     
      \hspace{-0.18 \textwidth}  \color{black}{(a)}
      \hspace{0.55\textwidth}  \color{black}{(b)}
         \hfill}
     \vspace{0.31\textwidth}    

\caption{Figure showing examples of the linear fit of the solar wind speed ($v_{sw}$) profile inside the MC. Panels a and b describe event 53 of Table \ref{S - Table A1} (WIND observations) and event 8 of Table \ref{Table A2} (Helios observations) respectively. The red lines in the plots denote the linear fit. $v_s$ and $v_e$ are the speeds at the MC start ($t_s$ ; blue dotted line) and at the MC end ($t_e$ ; black dotted line) respectively as obtained from the linear fit. We compute the MC expansion speeds (Eq~\ref{eq_vexp}) using $v_s$ and $v_e$. } 
   \label{Figure 0}
   \end{figure}

 For each of the ICMEs listed in Tables \ref{S - Table A1} and \ref{Table A2}, we compute the expansion speed of the MC cross-section using \citep{18NCSo}

\begin{equation}
\label{eq_vexp}
v_{exp} = \frac{1}{2}(v_s - v_e)
\end{equation}

where $v_s$ and $v_e$ are the speeds at the start and at the end of the MC boundary as obtained from a linear fit of the temporal profile of the solar wind speed (Figure \ref{Figure 0}). The expansion speed we determine is only for the MC part of the overall ICME. This is primarily because only MC boundaries are well determined for the Helios events; we retain the same scheme for the WIND events for consistency. 
However, the expansion speed need not be strictly zero in order for the static assumption (Eq~\ref{eq_static}) to be valid. If $|v_{exp}|$ is small in comparison with characteristic speeds such as the Alfv\'en speed, the static approximation can be considered to be valid. The WIND ICME website (\url{https://wind.nasa.gov/ICMEindex.php}) provides a thorough time profile of the Alfv\'en speed $v_{A}$, using which we compute the following ratio:

\begin{figure}    
   \centerline{\hspace*{0.015\textwidth}
               \includegraphics[width=0.515\textwidth , scale=0.6,width=8cm, height=6.0cm]{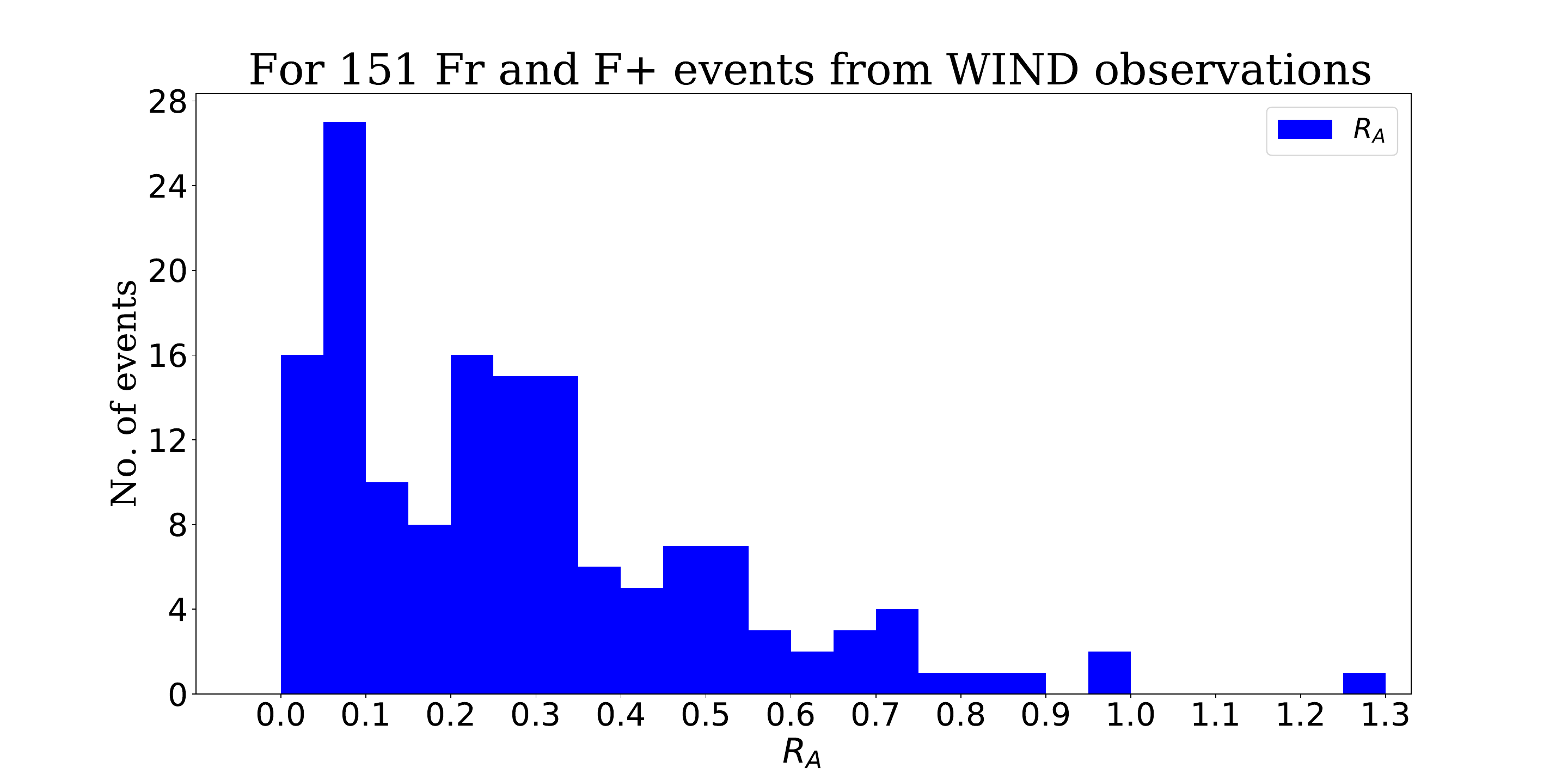}
               \hspace*{-0.03\textwidth}
               \includegraphics[width=0.515\textwidth , scale=0.6,width=8cm, height=6.0cm]{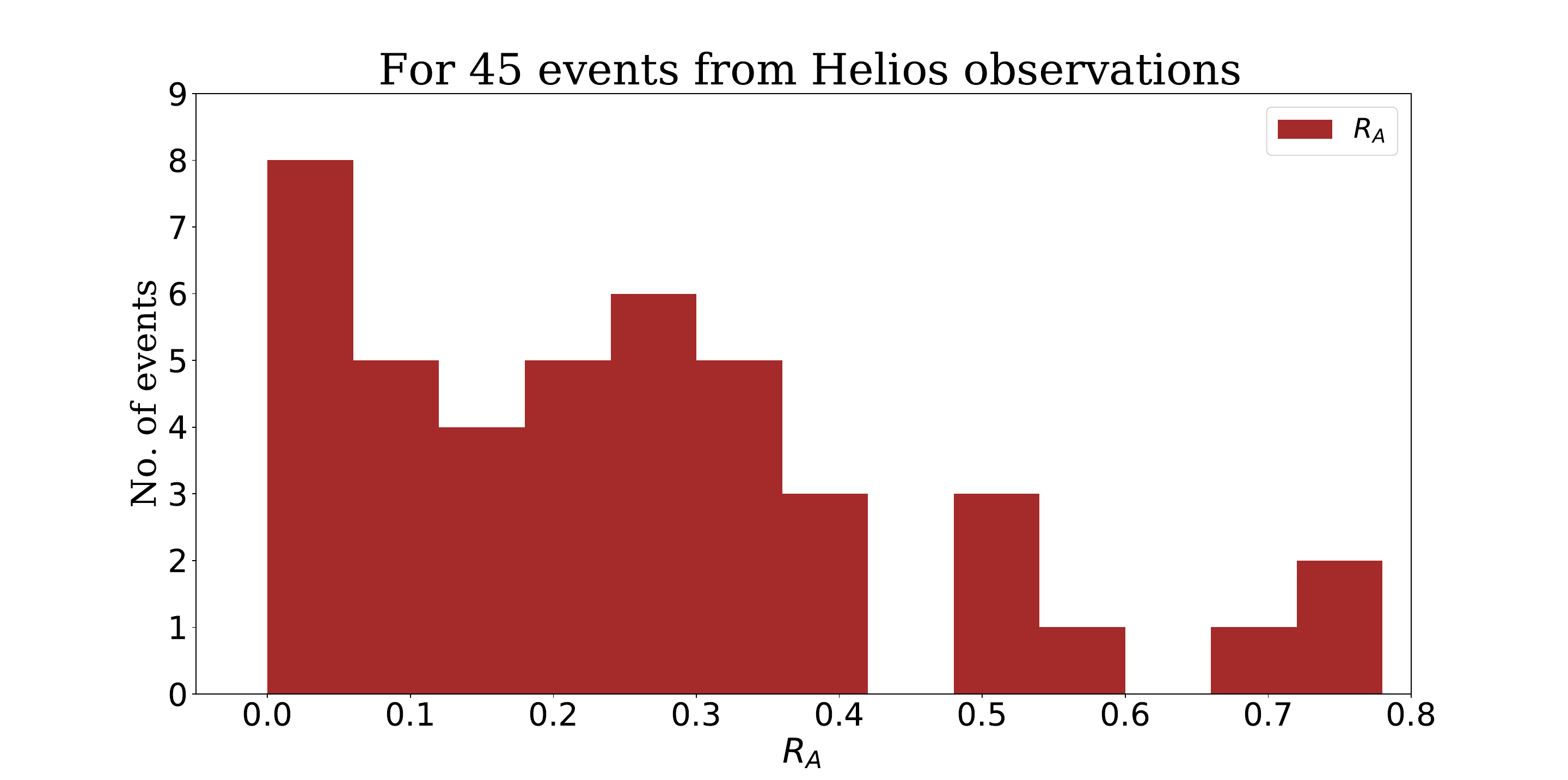}
              }
     \vspace{-0.35\textwidth}   
     \centerline{\Large \bf     
      \hspace{-0.18 \textwidth}  \color{black}{(a)}
      \hspace{0.55\textwidth}  \color{black}{(b)}
         \hfill}
     \vspace{0.31\textwidth}    

\caption{Histograms of $R_A$, the ratio of the MC expansion speed to the Alfv\'en speed (Eq~\ref{eq_Rs}) for the events listed in Tables \ref{S - Table A1} (panel a) and \ref{Table A2} (panel b). The mean, median and most probable values of the histogram in panel a are 0.29, 0.24 and 0.06 respectively. For panel b, the mean, median and most probable values are 0.28, 0.25 and 0.03 respectively.}
   \label{Figure 1}
   \end{figure}

\begin{equation}
R_{A} = \frac{|v_{exp}|}{\langle v_{A} \rangle} \, ,
\label{eq_Rs}
\end{equation}
where $\langle v_{A} \rangle$ is the average Alfv\'en speed inside the MC. 
Our findings for $R_{A}$ (Eq~\ref{eq_Rs}) for the WIND ICMEs (Table~\ref{S - Table A1}) and the Helios ICMEs (Table~\ref{Table A2}) are shown in Figure \ref{Figure 1}. The mean, median and most probable values for $R_{A}$ from the WIND events (panel a, Fig~\ref{Figure 1}) are 0.29, 0.24 and 0.06 respectively. The mean, median and most probable values of $R_{A}$ for the Helios events (panel b, Fig~\ref{Figure 1}) are 0.28, 0.25 and 0.03 respectively. With reference to point 1 in \S~\ref{S-Introduction}, we thus find that the expansion speeds of magnetic clouds are far smaller than the Alfv\'en speeds inside them. In other words, MC boundaries are static (to a fairly good extent) over an Alfv\'en crossing timescale.

\section{On ``MH'' in the MHD description}
\label{S - Lundquist}
The ``magnetohydro'' aspect of the MHD equations mandates that the medium is an infinitely conducting fluid. In turn, it implies that the magnetic diffusivity $\rightarrow 0$, which means that the magnetic field is frozen into the plasma. The Lundquist number, which is the ratio of the magnetic diffusion timescale to the Alfv\'en crossing timescale, is a measure of the goodness of this assumption. For an infinitely conducting plasma, the Lundquist number $\rightarrow \infty$. We check to see how well the MC plasma conforms to this expectation.
In order to evalute the Lundquist number in our situation, we need to know the conductivity of the MC plasma. The collisional timescales can be used to estimate the electrical conductivity.
In evaluating the electrical conductivity, we restrict ourselves to Coulomb collisions, and do not consider any anomalous effects such as scattering due to wave particle interactions. The timescale for electrons to relax to a Maxwellian distribution following a small perturbation is given by \citep{2013NRLHuba}

\begin{equation}
\label{tau_coll}
t_{ee} \approx 4 \times 10^4 \frac{{T_4}^{3/2}}{n\,ln(\Lambda)} \, ,
\end{equation}
where $T_4$ is the electron temperature in units of $10^4$ K, $n$ is the electron number density (assumed to be equal to the proton number density) in units of $cm^{-3}$ and $ln(\Lambda)$ is the Coulomb logarithm (taken to be 20 for our study). While Eq~\ref{tau_coll} gives the electron collisional relaxation timescale in units of second, the proton collisional relaxation timescale would be a factor of $\sqrt{m_{p}/m_{e}}$ larger and the electron-proton collisional equilibration timescale would be a factor of $m_{p}/m_{e}$ larger \citep{1994Sturrock}. We do not have direct access to electron temperature measurements; the WIND and Helios databases only provide measurements of proton temperature. Some authors \citep{1997RchJGR} suggest that the electron temperature exceeds the proton temperature by a factor $\gtrsim 2$ while some (e.g., \citet{1993OsJGR}) think that the factor can be between 7 and 10. Furthermore, the ICME electrons often have a thermal core and non-thermal wings \citep{2008NCJGRA}. For the sake of concreteness, we assume that the electron temperature is 10 times the proton temperature. Ohm's law for a magnetized plasma is generally written as

\begin{equation}
\mathbf{j} = \sigma_0 \mathbf{E}_\parallel + \sigma_P \mathbf{E}_\perp + \sigma_H (\mathbf{\hat{b}} \times \mathbf{E}) \, ,
\end{equation}
where $\mathbf{j}$ is the current density of the plasma, $\mathbf{\hat{b}}$ is the unit vector along $\mathbf{B}$, $\mathbf{E_\parallel}$ and $\mathbf{E_\perp}$ are the components of the electric field ($\mathbf{E}$) in the directions parallel and perpendicular to the $\mathbf{B}$ respectively.

\begin{figure}
\centerline{\includegraphics[width=0.9\textwidth,clip=]{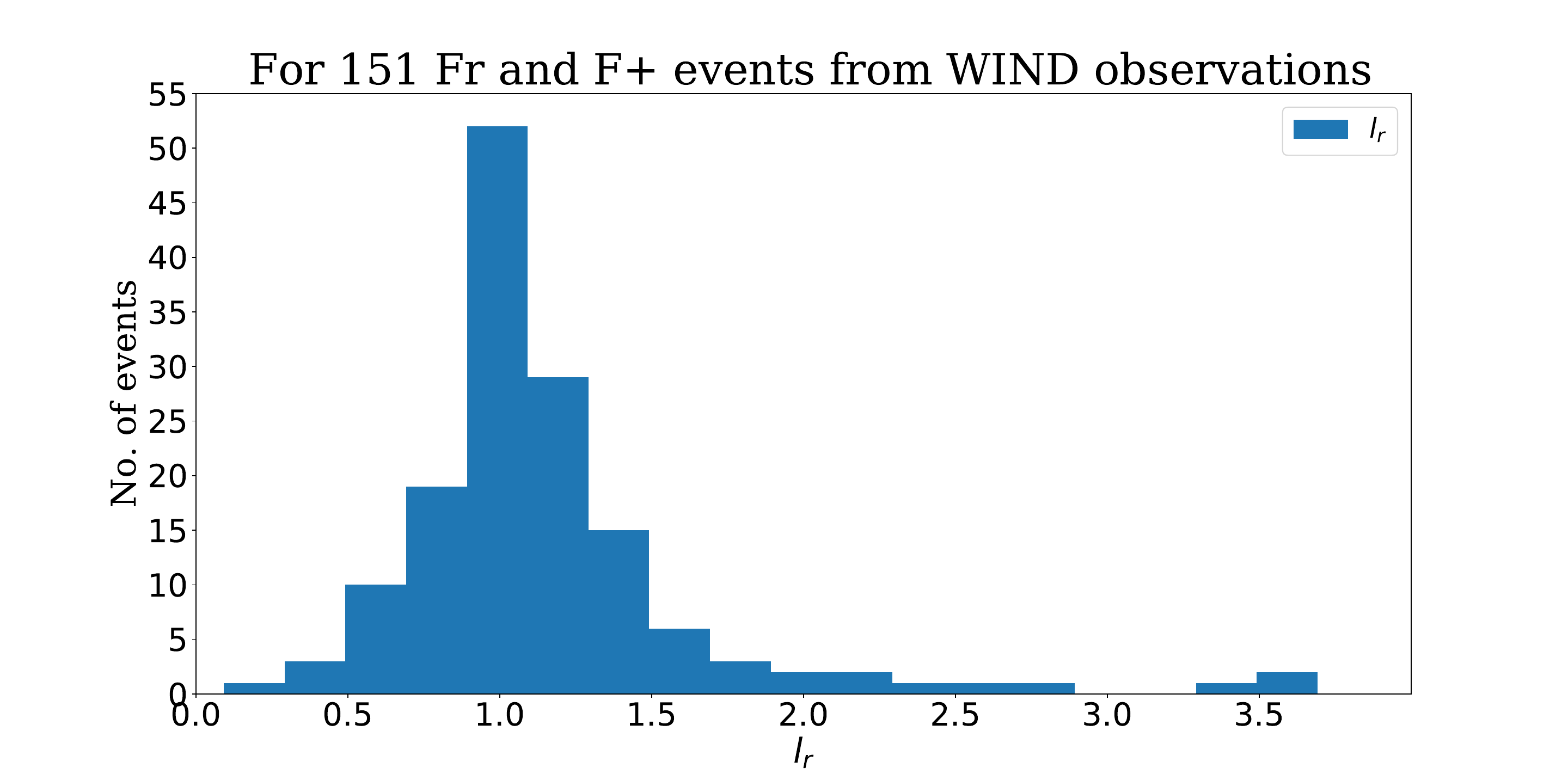}}
              
\caption{Histogram of $l_r$ (Equation \ref{eq: l_r}). The mean, median and the most probable values of $l_r$ are 1.22, 1.06 and 1.01 respectively.}
   \label{Figure l_r}              
\end{figure}

The different electrical conductivities are \citep{1994Sturrock}
{\begin{eqnarray}
\label{eq_sigmas}
\sigma_0 = \frac{n \, e^2\,t_{ee}}{m_e} \,\,\,\, {\rm isotropic \, conductivity} \\
\sigma_P = \sigma_0 \frac{{t_{ee}}^{-2}}{{t_{ee}}^{-2} + \Omega^2} \,\,\,\, {\rm Pederson \, conductivity} \\ 
\sigma_H = \sigma_0 \frac{t_{ee}^{-1} \Omega}{{t_{ee}}^{-2} + \Omega^2} \,\,\,\, {\rm Hall \, conductivity} \, ,
\end{eqnarray}}

where $e$ is the electron charge in cgs units, $m_e$ is the electron mass in g, $t_{ee}$ (Eq~\ref{tau_coll}) is the electron collisional timescale in second and $\Omega \equiv q |B|/m_{e} c$ (where $|B|$ is the magnitude of the magnetic field) is the electron gyrofrequency in Hz. The isotropic conductivity ($\sigma_{0}$) is the operative conductivity in an unmagnetized plasma; in a magnetized plasma, it denotes the conductivity along the magnetic field. The other two conductivities ($\sigma_{P}$ and $\sigma_{H}$) are the conductivities perpendicular to the magnetic field. The Lundquist number is usually defined only with regard to the magnetic diffusivity arising from the isotropic conductivity ($\sigma_{0}$). Accordingly, the isotropic magnetic diffusivity ($\eta_0$) and the corresponding diffusion timescale are given as:

{\begin{eqnarray}
\label{eq_etas}
\eta_0 = \frac{c^2}{4 \pi \sigma_0} \, , \,\,\, t_{\eta 0} = \frac{{L_{MC}}^2}{\eta_0}\,\,\,\,({\rm cgs\,\,units}) \, ,
\end{eqnarray}}
where $L_{MC}$ denotes the diameter of the magnetic cloud. For WIND events, we use
\begin{equation}
L_{MC} = 2 \times R_{cc} 
\label{eq: LMC_wind}
\end{equation}
where $R_{cc}$ is the MC radius fitted using the circular-cylindrical (cc) flux-rope model \citep{2002HidalgoJGRA, 2016NCApJ} and is available in the WIND ICME catalogue (\url{https://wind.nasa.gov/ICMEindex.php}). Since we do not have access to a quantity such as $R_{cc}$ for the 45 Helios events (Table \ref{Table A2}), we define $L_{MC}$ for these events as
\begin{equation}
L_{MC} \equiv \int_{t_s}^{t_e}   |\mathbf{v_{sw}}(t)| dt \, .
\label{eq:LMC} 
\end{equation} 
The quantity ${\mathbf v}_{sw}(t)$ denotes the solar wind velocity along the spacecraft trajectory, while $t_{s}$ and $t_{e}$ denote the start and end times of the MC respectively (Figure \ref{Figure 0}). By way of checking the reliability of $L_{MC}$ as defined in Equation \ref{eq:LMC}, we use a ratio

\begin{equation}
l_r = \frac{\int_{t_s}^{t_e}   |\mathbf{v_{sw}}(t)| dt}{2\times R_{cc}}
\label{eq: l_r}
\end{equation}
for the 151 WIND events (Figure \ref{Figure l_r}). The mean, median and the most probable values of the histogram displayed in Figure \ref{Figure l_r} are 1.22, 1.06 and 1.01 respectively. In other words, $l_r \approx 1$ and $L_{MC}$ as determined using Eq~\ref{eq:LMC} is quite close to that obtained by fitting a static flux rope model (Eq~\ref{eq: LMC_wind}). This justifies our use of Eq~\ref{eq:LMC} to calculate $L_{MC}$ for the Helios events (Table \ref{Table A2}).

The Lundquist number is defined as

\begin{figure}[h]  
   \centerline{\hspace*{0.015\textwidth}
               \includegraphics[width=0.515\textwidth , scale=0.6,width=8cm, height=6.0cm]{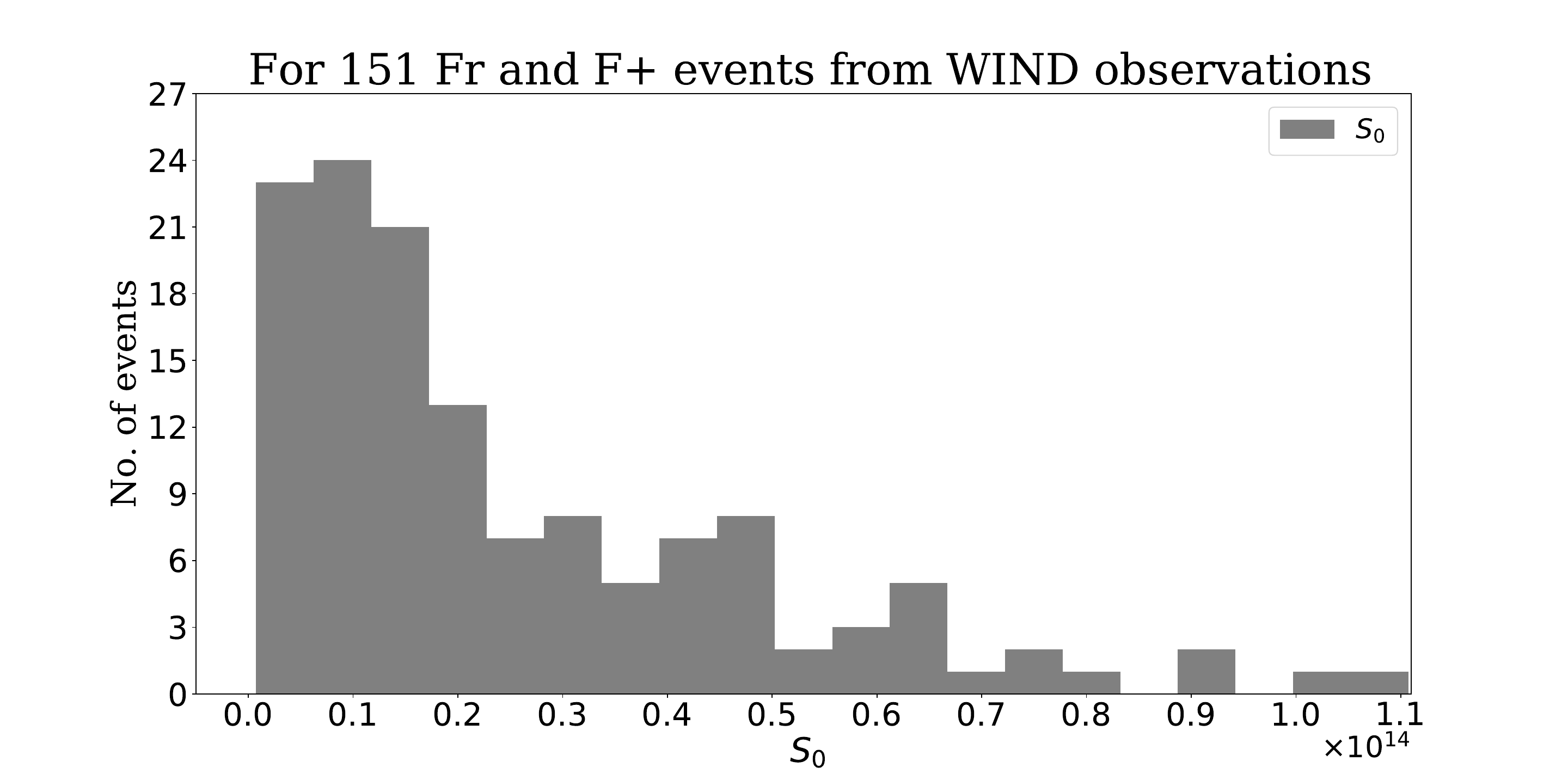}
               \hspace*{-0.03\textwidth}
               \includegraphics[width=0.515\textwidth , scale=0.6,width=8cm, height=6.0cm]{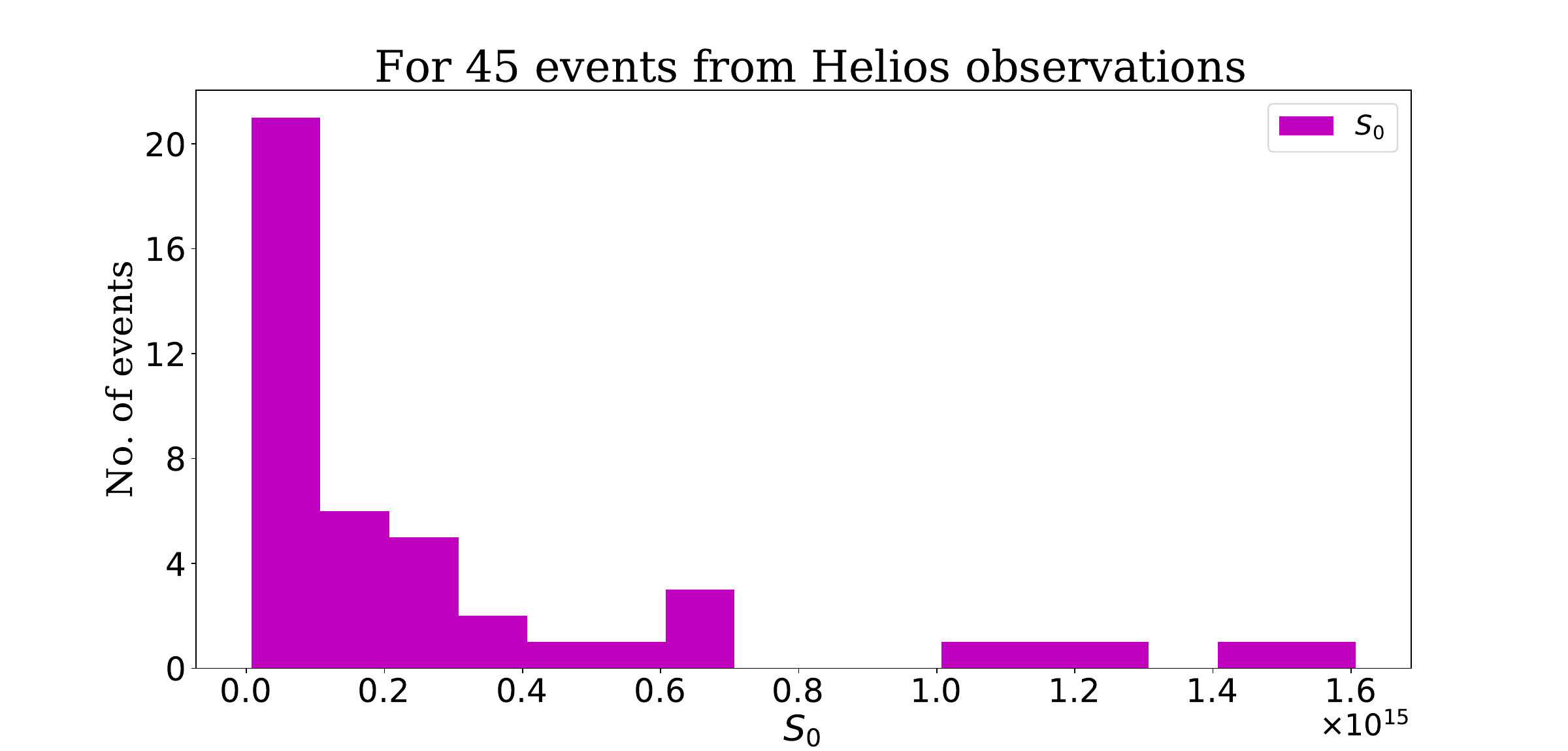}
              }
     \vspace{-0.35\textwidth}   
     \centerline{\Large \bf     
      \hspace{-0.18 \textwidth}  \color{black}{(a)}
      \hspace{0.55\textwidth}  \color{black}{(b)}
         \hfill}
     \vspace{0.31\textwidth}    

\caption{Histograms of the Lundquist number $S_0$ (Eq~\ref{eq_time_ratios1}) for the events listed in Table \ref{S - Table A1} (panel a) and Table \ref{Table A2} (panel b). The mean, median and the most probable values of the histogram in panel (a) are $9.8 \times 10^{13}$, $2 \times 10^{13}$ and $ 9\times 10^{12}$ respectively. The mean, median and most probable values for the histogram in panel (b) are $3.4 \times 10^{14}$, $1.4 \times 10^{14}$ and $4.3 \times 10^{13}$ respectively.}
   \label{Figure 3}
   \end{figure}

{\begin{eqnarray}
\label{eq_time_ratios1}
S_{0} = \bigg \langle \frac{ t_{\eta 0} }{ t_{A} }\bigg \rangle
\end{eqnarray}}
where the Alfv\'en timescale is $t_{A} \equiv L_{MC}/v_{A}$ and $\langle \, \, \rangle$ denotes an average inside the MC.
We evaluate the Lundquist number defined in Eq~\ref{eq_time_ratios1} on the set of events listed in Table \ref{S - Table A1} (observed by the WIND spacecraft) and Table \ref{Table A2} (observed by Helios 1 and 2 spacecrafts). The mean, median and most probable values of $S_0$ for the WIND events (panel a, Figure~\ref{Figure 3}) are $9.8 \times 10^{13}$, $2 \times 10^{13}$ and $ 9 \times 10^{12}$ respectively. The mean, median and most probable values of $S_0$ for the Helios events (panel b, Figure~\ref{Figure 3}) are $3.4 \times 10^{14}$, $1.4 \times 10^{14}$ and $4.3 \times 10^{13}$ respectively. As noted earlier, we have assumed the electron temperature to be 10 times the proton temperature for these calculations. If, instead, the electron temperature is taken to be equal to the proton temperature, the numbers on the $x$-axis of Figure~\ref{Figure 3} would need to be multiplied by 0.03. Consequently, the mean, median and most probable value would also be multiplied by 0.03. The main takeaway is that the magnetic diffusivity arising out of the isotropic conductivity is small, and the corresponding Lundquist number is therefore quite large. By comparison, the Lundquist number for laboratory Tokamak plasmas is $\gtrsim 10^{7}$ and that for plasmas in the magnetic reconnection experiment is $\lesssim 10^{3}$ \citep{1998JiPhRvL}. Going by this comparison, it is fair to conclude that MC plasmas adhere better (than laboratory plasmas) to the basic assumptions made in MHD. Since the electrons in ICMEs can be non-Maxwellian \citep{2008NCJGRA}, we note that transport coefficients such as the conductivity might be somewhat enhanced \citep{2021HusidicA&A}.

\section{Joule heating of MC plasma}
\label{S - Joule}
We now turn our attention to the dissipation rate implied by collisional conductivity calculated in \S~\ref{S - Lundquist}. In order to do this, we would need an estimate of the current density (${\mathbf j}$). The current density in ideal MHD is given by

\begin{equation}
\mathbf{j} = \frac{c}{4\pi}(\mathbf{\nabla \times B})
\label{eq: j_ideal}
\end{equation}

In order to calculate ${\mathbf j}$, we would need to know the curl of the magnetic field. We have a time series of the vector magnetic field along the line of intercept for each event in our dataset. We can approximate the spatial derivatives $\nabla$ by

\begin{equation}\mathbf{\nabla} \equiv \hat{x} \frac{\partial}{\partial x} + \hat{y} \frac{\partial}{\partial y}+ \hat{z} \frac{\partial}{\partial z} = \hat{x}\frac{1}{v_{sw\,x}} \frac{\partial}{\partial t} + \hat{y}\frac{1}{v_{sw\,y}} \frac{\partial}{\partial t} + \hat{z}\frac{1}{v_{sw\,z}} \frac{\partial}{\partial t} \, ,
\label{eq:nabla}
\end{equation}
where $v_{sw\,x}$ , $v_{sw\,y}$ and $v_{sw\,z}$ are the $x$ , $y$ and $z$ components of the solar wind velocity respectively. The idea employs the usual Taylor hypothesis, whereby we can write $\partial/\partial x \rightarrow v_{sw\,x}^{-1} \partial/\partial t$ and so on. The temporal derivatives ($\partial/\partial t$) are evaluated numerically from the time series data. Since we have access to all components of ${\mathbf B}$ and all components of ${\mathbf \nabla}$ (Eq~\ref{eq:nabla}), we can estimate $\mathbf{\nabla \times B}$ and consequently the current density ${\mathbf j}$ (Eq~\ref{eq: j_ideal}), at least along the line of spacecraft intercept. The directions ${\hat x}$, ${\hat y}$ and ${\hat z}$ refer to the cartesian coordinate system used for measuring the solar wind velocities and magnetic fields. For the WIND observations, this refers to the geo-centered solar ecliptic (GSE) co-ordinate system (\url{https://wind.nasa.gov/mfi_swe_plot.php}), whereas they refer to the the spacecraft-centered solar ecliptic (SSE) coordinate system for the Helios events \citep{2019DMJGRA}. The noise in the solar wind velocity measurements can introduce errors in our estimate of the current density; in particular, the $y$ and $z$ components of the solar wind velocity are found to be $\approx$ 1.4 to 1.9 times noisier than the $x$ component. On the other hand, the magnitude of $v_{sw\,x}$ is $\approx$ 100 times larger than that of the other two components. Errors can also be introduced via the numerical differentiation process. Notwithstanding these caveats, this method is a practical and reliable means of estimating the flux rope current, as we show herewith.

\begin{figure} [H]   
   \centerline{\hspace*{0.015\textwidth}
               \includegraphics[width=0.515\textwidth , scale=0.6,width=8cm, height=6.0cm]{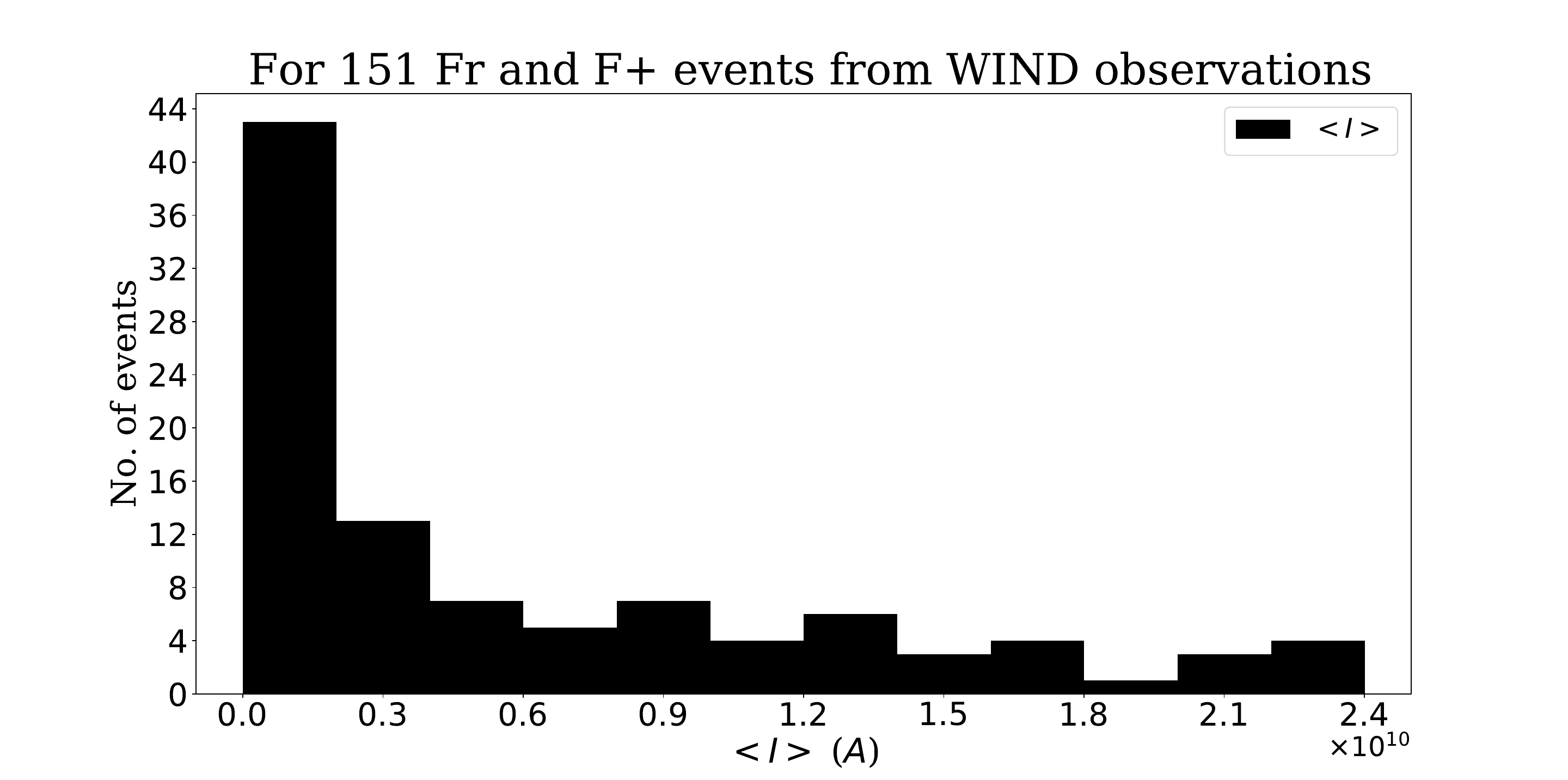}
               \hspace*{-0.03\textwidth}
               \includegraphics[width=0.515\textwidth , scale=0.6,width=8cm, height=6.0cm]{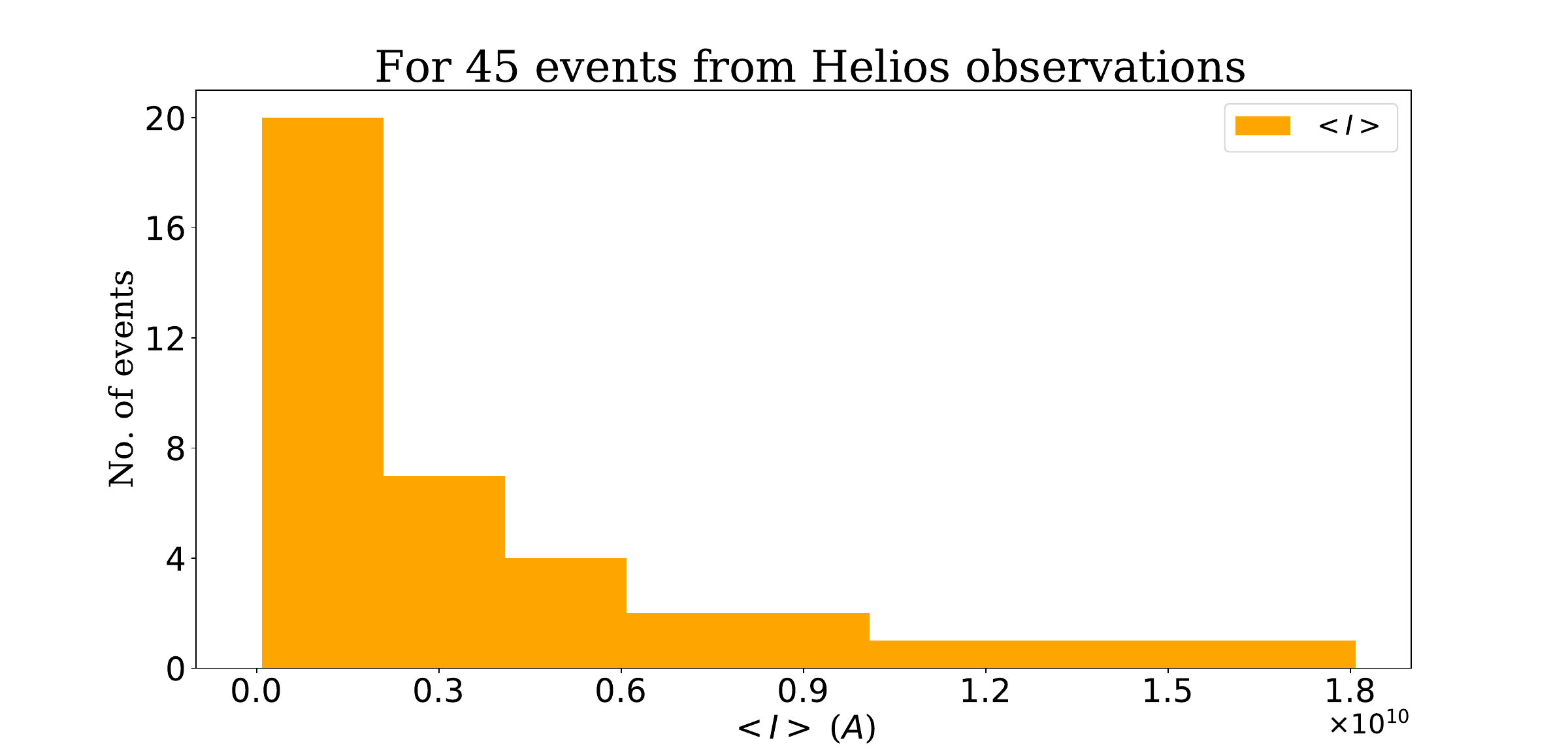}
              }
     \vspace{-0.35\textwidth}   
     \centerline{\Large \bf     
      \hspace{-0.18 \textwidth}  \color{black}{(a)}
      \hspace{0.55\textwidth}  \color{black}{(b)}
         \hfill}
     \vspace{0.31\textwidth}    

\caption{ Histograms describing the average current(Equation \ref{eq:avg current}) for the WIND (panel a) and Helios (panel b) events. The mean, median and the most probable values for the histogram of panel a are $10^{11}$A, $1.2 \times 10^{10}$A and $8 \times 10^8$A respectively. The mean, median and the most probable values for the Helios events are $5.8 \times 10^{10}$A, $3 \times 10^{9}$A and $10^9$A respectively.}
   \label{Figure I_avg}
   \end{figure}

An estimate of the average current carried by an MC can be calculated by

\begin{equation}
\label{eq:avg current}
\langle I \rangle = \langle \mathbf{j} \rangle \times (\pi/4) L_{MC}^{2}
\end{equation}

where the $\langle \, \rangle$ denotes averaging over the MC. 
The mean, median and most probable value of $\langle I \rangle$ for the WIND events are $ 10^{11}$A, $1.2 \times 10^{10}$A and $8 \times 10^8$A respectively (panel a of Figure \ref{Figure I_avg}), while the mean, median and most probable value of $\langle I \rangle$ for the Helios events are $5.8 \times 10^{10}$A, $3 \times 10^{9}$A and $10^9$A respectively (panel b of Figure \ref{Figure I_avg}). These numbers compare favorably with the estimate of $\approx 10^{9}$ A for the axial current carried by ICMEs near the Earth \citep{2019VrsnakApJ}. Axial currents for flux rope CMEs closer to the Sun (in the LASCO field of view) are somewhat larger; they are a few times $10^{10}$A \citep{2009PrasadApJ}. Our estimates of the axial current $\langle \, I \, \rangle$ are thus in reasonable agreement with other independent calculations, lending support to our estimates of the current density ${\mathbf j}$ using Eqs~\ref{eq: j_ideal} and \ref{eq:nabla}. 
The average rate of energy dissipation per unit mass due to Joule heating inside the MC can be written as

\begin{equation}
D_r =  {\bigg \langle} \frac{j^2}{\rho \sigma} {\bigg \rangle}
\label{eq: dissipation rate}
\end{equation}
where $\langle \, \, \rangle$ denotes averaging inside the MC, $\sigma \equiv \sigma_{0} + \sigma_{P} + \sigma_{H}$ (Eq~\ref{eq_sigmas}) and $\rho \equiv n(m_p + m_e)$ is the mass density, where $m_e$ and $m_p$ are the electron mass and proton mass respectively and $n$ is the proton number density (assumed to be equal to the electron number density). If the fluid comprising the MC was perfectly conducting (i.e., $\sigma \rightarrow \infty$), the Joule heating rate would $\rightarrow 0$.

\begin{figure} [H]   
   \centerline{\hspace*{0.015\textwidth}
               \includegraphics[width=0.515\textwidth , scale=0.6,width=8cm, height=6.0cm]{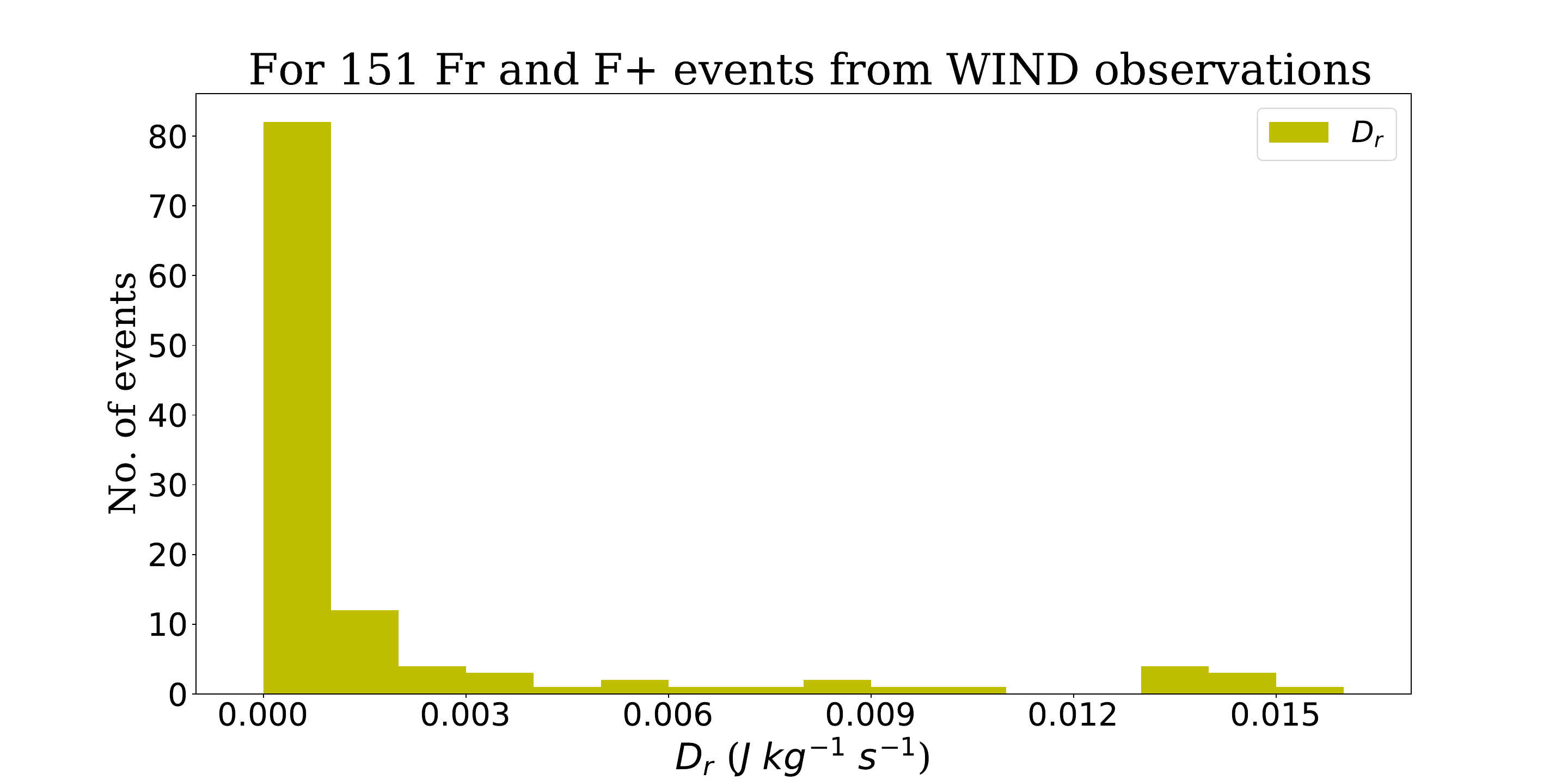}
               \hspace*{-0.03\textwidth}
               \includegraphics[width=0.515\textwidth , scale=0.6,width=8cm, height=6.0cm]{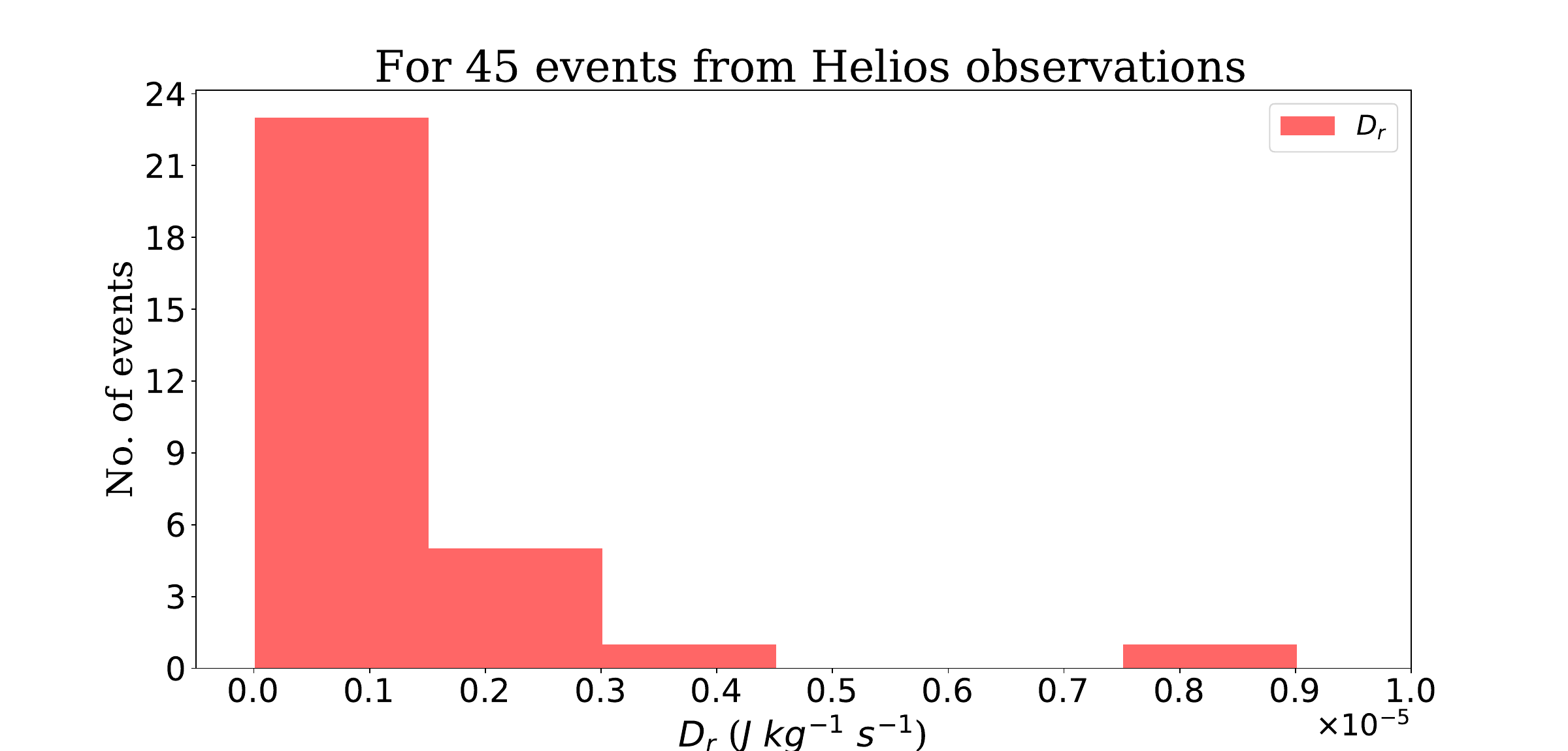}
              }
     \vspace{-0.35\textwidth}   
     \centerline{\Large \bf     
      \hspace{-0.18 \textwidth}  \color{black}{(a)}
      \hspace{0.55\textwidth}  \color{black}{(b)}
         \hfill}
     \vspace{0.31\textwidth}    

\caption{Histograms of the average Joule heating rate $D_r$ (Eq~\ref{eq: dissipation rate}). Panel a refers to the events listed in Table \ref{S - Table A1} while panel b refers the events listed in Table \ref{Table A2}. The mean, median and most probable values for the histogram in panel a are $1.1$ $Jkg^{-1}s^{-1}$, $6\times 10^{-4}$ $Jkg^{-1}s^{-1}$ and $8\times 10^{-6}$ $Jkg^{-1}s^{-1}$ respectively. The mean, median and most probable values for the histogram in panel b are $9.6 \times 10^{-3}$ $Jkg^{-1}s^{-1}$, $1.3 \times 10^{-6}$ $Jkg^{-1}s^{-1}$ and $2 \times 10^{-8}$ $Jkg^{-1}s^{-1}$ respectively.}
   \label{Figure 7}
   \end{figure}   
However, the collisional conductivities computed in \S~\ref{S - Lundquist} suggest a finite Joule heating rate $D_{r}$ (Eq~\ref{eq: dissipation rate}). We have plotted histograms of the Joule heating rate in units of ${\rm J\,kg^{-1}\,s^{-1}}$ in Figure \ref{Figure 7}. We now compare this heating rate with expectations from ICME models. If magnetic clouds expanded adiabatically, their proton temperatures would likely be as low as a few K near the earth \citep{1993CGGeoRL}. However, proton temperatures inside MCs are measured to be as high as $10^{5}$ K. This implies either that MCs remain magnetically connected to the solar corona with a very high thermal conductivity along the field \citep{1993CGGeoRL} or that there is substantial plasma heating inside MCs \citep{1996KRJGR}.  Assuming that there is some kind of a heating mechanism operating inside MCs, and using observations of the decrease in MC proton and alpha particle temperatures with heliocentric distance, \citet{2006LiuJGRA} estimate that the average required ICME heating rate at 1 AU is about 2550 ${\rm J\,kg^{-1}\, s^{-1}}$, of which 900 ${\rm J\,kg^{-1}\,s^{-1}}$ is accounted for by protons. By comparison, our estimates of the Joule heating rate inside MCs at near the Earth is much smaller. The mean, median and most probable value for the Joule heating rate for the WIND events (panel a, Figure~\ref{Figure 7}) are $1.1$, $6 \times 10^{-4}$ and $8 \times 10^{-6}$ ${\rm J\,kg^{-1}\,s^{-1}}$ respectively. The mean, median and most probable value for the Joule heating rate for the Helios events (panel b, Figure~\ref{Figure 7}) are $9\times 10^{-3}$, $1.3 \times 10^{-6}$ and $2 \times 10^{-8}$ ${\rm J\,kg^{-1}\,s^{-1}}$ respectively. In both cases, the mean value is biased by a few ($\approx 11$\% for WIND measurements and $\approx 6$\% for Helios measurements) events. The inadequacy of joule heating is primarily why turbulent dissipation and anomalous resistivity (which are clearly outside the purview of ideal MHD) are often invoked to account for proton heating (e.g., \citet{2006LiuJGRA, 1996VermaJGR}).

\section{Conclusions and final remarks}
\label{S - summary}

ICMEs are commonly modeled as static structures, where the usual assumptions of magnetohydrodynamics are valid, but the plasma bulk velocity is zero (Eq~\ref{eq_static}). While this approach is widely adopted, we are not aware of a systematic study using a large sample of ICMEs which examines the validity of these assumptions. The early study of \citet{1982KleinBurlagaJGR} mentions that magnetic cloud expansion speeds are expected to be $\approx 1/2$ the Alfv\'en speed in the {\em ambient} plasma, and some other papers \citep{2017Lugaz, 2020LugazApJ} support this statement. While it seems that the Alfv\'en speed refers to the medium surrounding the ICME in these papers,  \citet{2010RichCaneSoPh} refer to the Alfv\'en speed inside the ICME and state that ``the mean expansion speed is around half the Alfv\'en speed based on average ICME parameters''. Similarly, \citet{2006ZurbuchenSSRv} state ``The expansion speed $V_{EXP}$ is typically around half the Alfvén speed in the ICME''. It is also recognized that the ICME expansion speed should be lesser than the Alfv\'en speed in order for static ICME cross-section models such as ones using the Grad-Shafranov reconstruction technique to be valid \citep{2020FarrugiaJGRA}. In this paper, we use a sample of 151 ICMEs observed near the earth by the WIND spacecraft and 45 MCs observed between 0.3 and 1 AU by the Helios spacecrafts, to address the following questions (with a focus on the MC structures within the ICMEs):

\begin{enumerate}

\item ICMEs are observed to expand as they propagate, and yet their MC cross-sections are commonly modeled as static plasma columns. In what sense can the MC cross-sections be considered static? 
\item
Models of ICME and MC cross-sections assume the plasma to be a perfectly conducting fluid, which means that the Lundquist number (the ratio of the magnetic field diffusion timescale to the Alfv\'en timescale) should $\rightarrow \infty$. How large are the Lundquist numbers in the MC plasma?

\item
How does the Joule heating rate in the ICME plasma compare with the required heating rate (as mandated by the observed proton temperatures), and what does it imply about departures from ideal MHD assumptions?

\end{enumerate}

The first question is addressed in \S~\ref{S - static assumption}, where we find that MC expansion speeds are typically much smaller than Alfv\'en speeds (Figure \ref{Figure 1}). Although MC expansion speeds only indicate plasma velocities along the line of intercept by the spacecraft, our findings suggest that the timescales for signals to propagate from one end of the MC cross-section to the other $\ll$ those over which the cross-section expands, broadly supporting the notion of a static structure. The ratio $R_{A} \equiv |v_{exp}|/\langle v_{A} \rangle$ (Eq~\ref{eq_vexp}) has a mean of 0.29, a median of 0.24 and a most probable value of 0.06 for the WIND events in our sample, which are all observed at the position of WIND measurement. The mean, median and most probable values of $R_{A}$ for the Helios events are 0.28, 0.25 and 0.03 respectively. The Helios events are intercepted at various heliocentric distances, ranging from 0.3 to 1 AU. We note that the mean, median and most probable values of $R_A$ in all cases are below the frequently quoted value of 0.5.

The second question approaches the problem from a different perspective - one that is generally adopted in evaluating the suitability of the MHD approach in analyzing equilibrium configurations of plasma columns in laboratory settings. If the Lundquist number is large, it means that the plasma is effectively non-diffusive, and the magnetic field is frozen-in; a key assumption of MHD. This question is addressed in \S~\ref{S - Lundquist}. We use Coulomb collision timescales and the associated electrical conductivity to evaluate the Lundquist numbers, and find that they are large ($\gtrsim 10^{13}$) (Figure \ref{Figure 3}). Another way of appreciating the significance of large Lundquist numbers in this situation is as follows \citep{2003Boyd.Sanderson}: the MHD description technically mandates an infinite magnetic Reynolds number ($Re_m \equiv VL/\eta$ where $V$ is a representative macroscopic plasma velocity, $L$ is a representative macroscopic scale length and $\eta$ is the magnetic diffusivity). However, for a static plasma column, $V \rightarrow 0$ and so the magnetic Reynolds number ($Re_m$) must also $\rightarrow 0$. The only way of legitimizing an MHD description is to show that the magnetic diffusivity is very small; in other words, to show that the diffusion timescale is larger than any other timescale of interest (in this case, the Alfv\'en timescale). Large Lundquist numbers essentially ensure this.

The answer to the third question follows from \S~\ref{S - Joule}; the large Lundquist numbers imply that the resistive dissipation in the ICME plasma should be very small. They certainly are, especially in comparison with the plasma heating rate implied by the observed proton temperatures. Taken together, the answers to questions 1--3 generally offer strong support to the magnetohydrostatic modeling of ICME cross-sections.

However, there are some caveats to this seemingly robust conclusion. The standard MHD description precludes any heating/dissipation at all. We have calculated the Joule dissipation rate in MCs arising out of electron Coulomb collisions and found it to be several orders of magnitude lower that the heating requirements implied by the observed proton temperatures. Assuming efficient electron-proton energy exchange, this argues for additional electron heating that cannot be accounted for by Coulomb collisions. Of course, protons could also be preferentially heated by processes such as turbulent fluctuations. Observations of intense plasma heating in ICMEs at heliocentric distances of a few $R_{\odot}$ e.g., \citep{2011MurphyApJ, 2021Wilson} argue for localized electron heating (unlike the uniform Joule dissipation scenario we consider), perhaps in reconnection sites. The bulk ICME plasma could be heated via thermal conduction from these localized heating sites. Either way, it is clear that the enhanced energy dissipation required in ICMEs is one significant departure from the standard MHD description, which merits systematic study for a large sample of ICMEs. Conclusions from such a study could feed into dynamical models of Sun-Earth propagation and consequently on estimates of Earth arrival times. 


\section*{Acknowledgements}
\addcontentsline{toc}{section}{Acknowledgements}

We acknowledge the referee for constructive criticism that helped improve the manuscript. DB acknowledges a PhD studentship from the Indian Institute of Science Education and Research, Pune.


\appendix

\section{Data Table:}
\label{App - data table}

\subsection{Events from the WIND ICME catalogue}
\label{App - wind table}

 
\begin{table}[H]

\caption{
The list of the 151 WIND ICME events we use in this study. The arrival date and time of the ICME at the position of WIND measurement and the arrival and departure dates \& times of the associated magnetic clouds (MCs) are taken from WIND ICME catalogue (\url{https://wind.nasa.gov/ICMEindex.php}). The 15 events marked with and asterisk (*) coincide with the near earth counterparts of 15 CMEs listed in \citet{17NSoPh}. }  
  \begin{center}
  
  \begin{tabular}{cccclccc}
  
   \hline
   CME        & CME Arrival date & MC start  & MC end       & Flux rope \\
   event      & and time[UT]   &  date and   & date and     & type   \\
    number    &  (1AU)         &  time[UT]   &   time[UT]   &      \\
    \hline
    \hline
 
 1    &  1995 03 04 , 00:36 & 1995 03 04 , 11:23 & 1995 03 05 , 03:06 & Fr \\
 2    &  1995 04 03 , 06:43 & 1995 04 03 , 12:45 & 1995 04 04 , 13:25 & F+ \\
 3    &  2010 06 30 , 09:21 & 1995 06 30 , 14:23 & 1995 07 02 , 16:47 & Fr \\
 4    &  1995 08 22 , 12:56 & 1995 08 22 , 22:19 & 1995 08 23 , 18:43 & Fr \\
 5    &  1995 09 26 , 15:57 & 1995 09 27 , 03:36 & 1995 09 27 , 21:21 & Fr \\
 6    &  1995 10 18 , 10:40 & 1995 10 18 , 19:11 & 1995 10 20 , 02:23 & Fr \\
 7    &  1996 02 15 , 15:07 & 1996 02 15 , 15:07 & 1996 02 16 , 08:59 & F+ \\
 8    &  1996 04 04 , 11:59 & 1996 04 04 , 11:59 & 1996 04 04 , 21:36 & Fr \\ 
 9    &  1996 05 16 , 22:47 & 1996 05 17 , 01:36 & 1996 05 17 , 11:58 & F+ \\
 10   &  1996 05 27 , 14:45 & 1996 05 27 , 14:45 & 1996 05 29 , 02:22 & Fr \\
 11   &  1996 07 01 , 13:05 & 1996 07 01 , 17:16 & 1996 07 02 , 10:17 & Fr \\
 12   &  1996 08 07 , 08:23 & 1996 08 07 , 11:59 & 1996 08 08 , 13:12 & Fr \\
 13   &  1996 12 24 , 01:26 & 1996 12 24 , 03:07 & 1996 12 25 , 11:44 & F+ \\ 
 14   &  1997 01 10 , 00:52 & 1997 01 10 , 04:47 & 1997 01 11 , 03:36 & F+ \\
 15   &  1997 04 10 , 17:02 & 1997 04 11 , 05:45 & 1997 04 11 , 19:10 & Fr \\
 16   &  1997 04 21 , 10:11 & 1997 04 21 , 11:59 & 1997 04 23 , 07:11 & F+ \\
 17   &  1997 05 15 , 01:15 & 1997 05 15 , 10:00 & 1997 05 16 , 02:37 & F+ \\
 18   &  1997 05 26 , 09:09 & 1997 05 26 , 15:35 & 1997 05 28 , 00:00 & Fr \\
 19   &  1997 06 08 , 15:43 & 1997 06 09 , 06:18 & 1997 06 09 , 23:01 & Fr \\
 20   &  1997 06 19 , 00:00 & 1997 06 19 , 05:31 & 1997 06 20 , 22:29 & Fr \\
 21   &  1997 07 15 , 03:10 & 1997 07 15 , 06:48 & 1997 07 16 , 11:16 & F+ \\
 22   &  1997 08 03 , 10:10 & 1997 08 03 , 13:55 & 1997 08 04 , 02:23 & Fr \\
 23   &  1997 08 17 , 01:56 & 1997 08 17 , 06:33 & 1997 08 17 , 20:09 & Fr \\
 24   &  1997 09 02 , 22:40 & 1997 09 03 , 08:38 & 1997 09 03 , 20:59 & Fr \\
 25   &  1997 09 18 , 00:30 & 1997 09 18 , 04:07 & 1997 09 19 , 23:59 & F+ \\
 26   &  1997 10 01 , 11:45 & 1997 10 01 , 17:08 & 1997 10 02 , 23:15 & Fr \\
 27   &  1997 10 10 , 03:08 & 1997 10 10 , 15:33 & 1997 10 11 , 22:00 & F+ \\
 28   &  1997 11 06 , 22:25 & 1997 11 07 , 06:00 & 1997 11 08 , 22:46 & F+ \\
 29   &  1997 11 22 , 09:12 & 1997 11 22 , 17:31 & 1997 11 23 , 18:43 & F+ \\
 30   &  1997 12 30 , 01:13 & 1997 12 30 , 09:35 & 1997 12 31 , 08:51 & Fr \\
 31   &  1998 01 06 , 13:29 & 1998 01 07 , 02:23 & 1998 01 08 , 07:54 & F+ \\
 32   &  1998 01 28 , 16:04 & 1998 01 29 , 13:12 & 1998 01 31 , 00:00 & F+ \\
 33   &  1998 03 25 , 10:48 & 1998 03 25 , 14:23 & 1998 03 26 , 08:57 & Fr \\
 34   &  1998 03 31 , 07:11 & 1998 03 31 , 11:59 & 1998 04 01 , 16:18 & Fr \\
 35   &  1998 05 01 , 21:21 & 1998 05 02 , 11:31 & 1998 05 03 , 16:47 & Fr \\
 36   &  1998 06 02 , 10:28 & 1998 06 02 , 10:28 & 1998 06 02 , 09:16 & Fr \\
 37   &  1998 06 24 , 10:47 & 1998 06 24 , 13:26 & 1998 06 25 , 22:33 & F+ \\
 38   &  1998 07 10 , 22:36 & 1998 07 10 , 22:36 & 1998 07 12 , 21:34 & F+ \\
 39   &  1998 08 19 , 18:40 & 1998 08 20 , 08:38 & 1998 08 21 , 20:09 & F+ \\
 40   &  1998 10 18 , 19:30 & 1998 10 19 , 04:19 & 1998 10 20 , 07:11 & F+ \\
 41   &  1999 02 11 , 17:41 & 1999 02 11 , 17:41 & 1999 02 12 , 03:35 & Fr \\
 42   &  1999 07 02 , 00:27 & 1999 07 03 , 08:09 & 1999 07 05 , 13:13 & Fr \\
 43   &  1999 09 21 , 18:57 & 1999 09 21 , 18:57 & 1999 09 22 , 11:31 & Fr \\
 44   &  2000 02 11 , 23:34 & 2000 02 12 , 12:20 & 2000 02 13 , 00:35 & Fr \\
 45   &  2000 03 01 , 01:58 & 2000 03 01 , 03:21 & 2000 03 02 , 03:07 & Fr \\
 \hline
 
 \end{tabular}
 \end{center}
 \end{table}
 

 \begin{table}[H]
\begin{center}
 \begin{tabular}{cccclccc}
 CME        & CME Arrival date & MC start  & MC end       & Flux rope \\
   event      & and time[UT]   &  date and   & date and     & type   \\
    number    &  (1AU)         &  time[UT]   &   time[UT]   &      \\
    \hline
    \hline

 
 46   &  2000 07 01 , 07:12 & 2000 07 01 , 07:12 & 2000 07 02 , 03:34 & Fr \\
 47   &  2000 07 11 , 22:35 & 2000 07 11 , 22:35 & 2000 07 13 , 04:33 & Fr \\
 48   &  2000 07 28 , 06:38 & 2000 07 28 , 14:24 & 2000 07 29 , 10:06 & F+ \\
 49   &  2000 09 02 , 23:16 & 2000 09 02 , 23:16 & 2000 09 03 , 22:32 & Fr \\
 50   &  2000 10 03 , 01:02 & 2000 10 03 , 09:36 & 2000 10 05 , 03:34 & F+ \\
 51   &  2000 10 12 , 22:33 & 2000 10 13 , 18:24 & 2000 10 14 , 19:12 & Fr \\
 52   &  2000 11 06 , 09:30 & 2000 11 06 , 23:05 & 2000 11 07 , 18:05 & Fr \\
 53   &  2000 11 26 , 11:43 & 2000 11 27 , 09:30 & 2000 11 28 , 09:36 & Fr \\
 54   &  2001 04 21 , 15:29 & 2001 04 22 , 00:28 & 2001 04 23 , 01:11 & Fr \\
 55   &  2001 10 21 , 16:39 & 2001 10 22 , 01:17 & 2001 10 23 , 00:47 & Fr \\
 56   &  2001 11 24 , 05:51 & 2001 11 24 , 15:47 & 2001 11 25 , 13:17 & Fr \\
 57   &  2001 12 29 , 05:16 & 2001 12 30 , 03:24 & 2001 12 30 , 19:10 & Fr \\
 58   &  2002 02 28 , 05:06 & 2002 02 28 , 19:11 & 2002 03 01 , 23:15 & Fr \\
 59   &  2002 03 18 , 13:14 & 2002 03 19 , 06:14 & 2002 03 20 , 15:36 & Fr \\
 60   &  2002 03 23 , 11:24 & 2002 03 24 , 13:11 & 2002 03 25 , 21:36 & Fr \\
 61   &  2002 04 17 , 11:01 & 2002 04 17 , 21:36 & 2002 04 19 , 08:22 & F+ \\
 62   &  2002 07 17 , 15:56 & 2002 07 18 , 13:26 & 2002 07 19 , 09:35 & Fr \\
 63   &  2002 08 18 , 18:40 & 2002 08 19 , 19:12 & 2002 08 21 , 13:25 & Fr \\
 64   &  2002 08 26 , 11:16 & 2002 08 26 , 14:23 & 2002 08 27 , 10:47 & Fr \\
 65   &  2002 09 30 , 07:54 & 2002 09 30 , 22:04 & 2002 10 01 , 20:08 & F+ \\
 66   &  2002 12 21 , 03:21 & 2002 12 21 , 10:20 & 2002 12 22 , 15:36 & Fr \\
 67   &  2003 01 26 , 21:43 & 2003 01 27 , 01:40 & 2003 01 27 , 16:04 & Fr \\
 68   &  2003 02 01 , 13:06 & 2003 02 02 , 19:11 & 2003 02 03 , 09:35 & Fr \\
 69   &  2003 03 20 , 04:30 & 2003 03 20 , 11:54 & 2003 03 20 , 22:22 & Fr \\
 70   &  2003 06 16 , 22:33 & 2003 06 16 , 17:48 & 2003 06 18 , 08:18 & Fr \\
 71   &  2003 08 04 , 20:23 & 2003 08 05 , 01:10 & 2003 08 06 , 02:23 & Fr \\
 72   &  2003 11 20 , 08:35 & 2003 11 20 , 11:31 & 2003 11 21 , 01:40 & Fr \\
 73   &  2004 04 03 , 09:55 & 2004 04 04 , 01:11 & 2004 04 05 , 19:11 & F+ \\
 
 74   &  2005 05 15 , 02:10 & 2005 05 15 , 05:31 & 2005 05 16 , 22:47 & F+ \\
 75   &  2005 05 20 , 04:47 & 2005 05 20 , 09:35 & 2005 05 22 , 02:23 & F+ \\
 76   &  2005 07 17 , 14:52 & 2005 07 17 , 14:52 & 2005 07 18 , 05:59 & Fr \\
 77   &  2005 10 31 , 02:23 & 2005 10 31 , 02:23 & 2005 10 31 , 18:42 & Fr \\
 78   &  2006 02 05 , 18:14 & 2006 02 05 , 20:23 & 2006 02 06 , 11:59 & F+ \\
 79   &  2006 09 30 , 02:52 & 2006 09 30 , 08:23 & 2006 09 30 , 22:03 & F+ \\
 80   &  2006 11 18 , 07:11 & 2006 11 18 , 07:11 & 2006 11 20 , 04:47 & Fr \\
 81   &  2007 05 21 , 22:40 & 2007 05 21 , 22:45 & 2007 05 22 , 13:25 & Fr \\
 82   &  2007 06 08 , 05:45 & 2007 06 08 , 05:45 & 2007 06 09 , 05:15 & Fr \\
 83   &  2007 11 19 , 17:22 & 2007 11 20 , 00:33 & 2007 11 20 , 11:31 & Fr \\
 84   &  2008 05 23 , 01:12 & 2008 05 23 , 01:12 & 2008 05 23 , 10:46 & F+ \\
 85   &  2008 09 03 , 16:33 & 2008 09 03 , 16:33 & 2008 09 04 , 03:49 & F+ \\
 86   &  2008 09 17 , 00:43 & 2008 09 17 , 03:57 & 2008 09 18 , 08:09 & Fr \\
 87   &  2008 12 04 , 11:59 & 2008 12 04 , 16:47 & 2008 12 05 , 10:47 & Fr \\
 88   &  2008 12 17 , 03:35 & 2008 12 17 , 03:35 & 2008 12 17 , 15:35 & Fr \\
 89   &  2009 02 03 , 19:21 & 2009 02 03 , 01:12 & 2009 02 04 , 19:40 & F+ \\
 90   &  2009 03 11 , 22:04 & 2009 03 12 , 01:12 & 2009 03 13 , 01:40 & F+ \\
 91   &  2009 04 22 , 11:16 & 2009 04 22 , 14:09 & 2009 04 22 , 20:37 & Fr \\
 92   &  2009 06 03 , 13:40 & 2009 06 03 , 20:52 & 2009 06 05 , 05:31 & Fr \\
 93   &  2009 06 27 , 11:02 & 2009 06 27 , 17:59 & 2009 06 28 , 20:24 & F+ \\
 94   &  2009 07 21 , 02:53 & 2009 07 21 , 04:48 & 2009 07 22 , 03:36 & Fr \\
 95   &  2009 09 10 , 10:19 & 2009 09 10 , 10:19 & 2009 09 10 , 19:26 & Fr \\
 \hline
 
 \end{tabular}
 \end{center}
 \end{table}

 
\begin{table}[H]
\begin{center}
 \begin{tabular}{cccclccc}
 CME        & CME Arrival date & MC start  & MC end       & Flux rope \\
   event      & and time[UT]   &  date and   & date and     & type   \\
    number    &  (1AU)         &  time[UT]   &   time[UT]   &      \\
    \hline
    \hline

 96   &  2009 09 30 , 00:44 & 2009 09 30 , 06:59 & 2009 09 30 , 19:11 & Fr \\
 97   &  2009 10 29 , 01:26 & 2009 10 29 , 01:26 & 2009 10 29 , 23:45 & F+ \\
 98   &  2009 11 14 , 10:47 & 2009 11 14 , 10:47 & 2009 11 15 , 11:45 & Fr \\
 99   &  2009 12 12 , 04:47 & 2009 12 12 , 19:26 & 2009 12 14 , 04:47 & Fr \\
 100   &  2010 01 01 , 22:04 & 2010 01 02 , 00:14 & 2010 01 03 , 09:06 & Fr \\
 101   &  2010 02 07 , 18:04 & 2010 02 07 , 19:11 & 2010 02 09 , 05:42 & Fr \\
 102*   &  2010 03 23 , 22:29 & 2010 03 23 , 22:23 & 2010 03 24 , 15:36 & Fr \\
 103*   &  2010 04 05 , 07:55 & 2010 04 05 , 11:59 & 2010 04 06 , 16:48 & Fr \\
 104*   &  2010 04 11 , 12:20 & 2010 04 11 , 21:36 & 2010 04 12 , 14:12 & Fr \\
 105   &  2010 05 28 , 01:55 & 2010 05 29 , 19:12 & 2010 05 29 , 17:58 & Fr \\
 106*   &  2010 06 21 , 03:35 & 2010 06 21 , 06:28 & 2010 06 22 , 12:43 & Fr \\
 107*   &  2010 09 15 , 02:24 & 2010 09 15 , 02:24 & 2010 09 16 , 11:58 & Fr \\
 108*   &  2010 10 31 , 02:09 & 2010 10 30 , 05:16 & 2010 11 01 , 20:38 & Fr \\
 109   &  2010 12 19 , 00:35 & 2010 12 19 , 22:33 & 2010 12 20 , 22:14 & F+ \\
 110   &  2011 01 24 , 06:43 & 2011 01 24 , 10:33 & 2011 01 25 , 22:04 & F+ \\
 111*   &  2011 03 29 , 15:12 & 2011 03 29 , 23:59 & 2011 04 01 , 14:52 & Fr \\
 112   &  2011 05 28 , 00:14 & 2011 05 28 , 05:31 & 2011 05 28 , 22:47 & F+ \\
 113   &  2011 06 04 , 20:06 & 2011 06 05 , 01:12 & 2011 06 05 , 18:13 & Fr \\
 114   &  2011 07 03 , 19:12 & 2011 07 03 , 19:12 & 2011 07 04 , 19:12 & Fr \\
 115*   &  2011 09 17 , 02:57 & 2011 09 17 , 15:35 & 2011 09 18 , 21:07 & Fr \\
 116   &  2012 02 14 , 07:11 & 2012 02 14 , 20:52 & 2012 02 16 , 04:47 & Fr \\
 117   &  2012 04 05 , 14:23 & 2012 04 05 , 19:41 & 2012 04 06 , 21:36 & Fr \\
 118   &  2012 05 03 , 00:59 & 2012 05 04 , 03:36 & 2012 05 05 , 11:22 & Fr \\
 119   &  2012 05 16 , 12:28 & 2012 05 16 , 16:04 & 2012 05 18 , 02:11 & Fr \\
 120   &  2012 06 11 , 02:52 & 2012 06 11 , 11:31 & 2012 06 12 , 05:16 & Fr \\
 121*   &  2012 06 16 , 09:03 & 2012 06 16 , 22:01 & 2012 06 17 , 11:23 & F+ \\
 122*   &  2012 07 14 , 17:39 & 2012 07 15 , 06:14 & 2012 07 17 , 03:22 & Fr \\
 123   &  2012 08 12 , 12:37 & 2012 08 12 , 19:12 & 2012 08 13 , 05:01 & Fr \\
 124   &  2012 08 18 , 03:25 & 2012 08 18 , 19:12 & 2012 08 19 , 08:22 & Fr \\
 125*   &  2012 10 08 , 04:12 & 2012 10 08 , 15:50 & 2012 10 09 , 17:17 & Fr \\
 126   &  2012 10 12 , 08:09 & 2012 10 12 , 18:09 & 2012 10 13 , 09:14 & Fr \\
 127*   &  2012 10 31 , 14:28 & 2012 10 31 , 23:35 & 2012 11 02 , 05:21 & F+ \\
 128*   &  2012 11 12 , 22:12 & 2012 11 13 , 08:23 & 2012 11 14 , 08:09 & F+ \\
 129*   &  2013 03 17 , 05:21 & 2013 03 17 , 14:09 & 2013 03 19 , 16:04 & Fr \\
 130*   &  2013 04 13 , 22:13 & 2013 04 14 , 17:02 & 2013 04 17 , 05:30 & F+ \\
 131   &  2013 04 30 , 08:52 & 2013 04 30 , 12:00 & 2013 05 01 , 07:12 & Fr \\
 132   &  2013 05 14 , 02:23 & 2013 05 14 , 06:00 & 2013 05 15 , 06:28 & Fr \\
 133   &  2013 06 06 , 02:09 & 2013 06 06 , 14:23 & 2013 06 08 , 00:00 & F+ \\
 134   &  2013 06 27 , 13:51 & 2013 06 28 , 02:23 & 2013 06 29 , 11:59 & Fr \\
 135   &  2013 09 01 , 06:14 & 2013 09 01 , 13:55 & 2013 09 02 , 01:56 & Fr \\
 136   &  2013 10 30 , 18:14 & 2013 10 30 , 18:14 & 2013 10 31 , 05:30 & Fr \\
 137   &  2013 11 08 , 21:07 & 2013 11 08 , 23:59 & 2013 11 09 , 06:14 & Fr \\
 138   &  2013 11 23 , 00:14 & 2013 11 23 , 04:47 & 2013 11 23 , 15:35 & Fr \\
 139   &  2013 12 14 , 16:47 & 2013 12 15 , 16:47 & 2013 12 16 , 05:30 & Fr \\
 140   &  2013 12 24 , 20:36 & 2013 12 25 , 04:47 & 2013 12 25 , 17:59 & F+ \\
 
 \hline
 
 \end{tabular}
 \end{center}
 \end{table}
 

  \begin{table}[H]
\begin{center}
 \begin{tabular}{cccclccc}
 CME        & CME Arrival date & MC start  & MC end       & Flux rope \\
   event      & and time[UT]   &  date and   & date and     & type   \\
    number    &  (1AU)         &  time[UT]   &   time[UT]   &      \\
    \hline
    \hline

 
 141   &  2014 04 05 , 09:58 & 2014 04 05 , 22:18 & 2014 04 07 , 14:24 & Fr \\
 142   &  2014 04 11 , 06:57 & 2014 04 11 , 06:57 & 2014 04 12 , 20:52 & F+ \\
 143   &  2014 04 14 , 10:20 & 2014 04 21 , 07:41 & 2014 04 22 , 06:12 & Fr \\
 144   &  2014 04 29 , 19:11 & 2014 04 29 , 19:11 & 2014 04 30 , 16:33 & Fr \\
 145   &  2014 06 29 , 04:47 & 2014 06 29 , 20:53 & 2014 06 30 , 11:15 & Fr \\
 146   &  2014 08 19 , 05:49 & 2014 08 19 , 17:59 & 2014 08 21 , 19:09 & F+ \\
 147   &  2014 08 26 , 02:40 & 2014 08 27 , 03:07 & 2014 08 27 , 21:49 & Fr \\
 148   &  2015 01 07 , 05:38 & 2015 01 07 , 06:28 & 2015 01 07 , 21:07 & F+ \\
 149   &  2015 09 07 , 13:05 & 2015 09 07 , 23:31 & 2015 09 09 , 14:52 & F+ \\
 150   &  2015 10 06 , 21:35 & 2015 10 06 , 21:35 & 2015 10 07 , 10:03 & Fr \\
 151   &  2015 12 19 , 15:35 & 2015 12 20 , 13:40 & 2015 12 21 , 23:02 & Fr \\
 \hline
 
  \end{tabular}
  
  \label{S - Table A1}
  \end{center}
\end{table}      


 \subsection{Events from the Helios Observation}
\label{Table A2}


\begin{table}[H]

\caption{The list of 45 magnetic clouds (MCs) shortlisted by \citet{1998BthAnGeo} using {\em in-situ} data from the Helios 1 and 2 spacecrafts during December 1974--July 1981. The columns show the year, time duration, heliocentric distance and the spacecraft by which the MC was observed. 
}  
  \begin{center}
  \begin{tabular}{cccccccc}
  
   \hline
   Magnetic Cloud   & Year     & \multicolumn{2}{c}{Time duration} &  Heliocentric distance       & Observed \\
   serial  number &     & Days    &  Hours   & (AU)  &  by  \\
    \hline
    \hline

1 &  1975   &  7  &  00 -- 10  &  0.92  & Helios 1 \\ 

2 &         &  63,64  &  16 -- 05  &  0.39  & Helios 1 \\ 

3 &         &  92  &  06 -- 16  &  0.48  & Helios 1 \\ 

4 &         &  313  &  03 -- 18  &  0.81  & Helios 1 \\ 

5 &         &  321  &  06 -- 18  &  0.87  & Helios 1 \\ 

6 &  1976   &  90  &  09 -- 21  &  0.47  & Helios 2 \\ 

7 &         &  187  &  03 -- 21  &  0.98  & Helios 1 \\ 

8 &  1977   &  29,30  &  10 -- 10  &  0.95  & Helios 1 \\ 

9 &         &  76  &  05 -- 20  &  0.71  & Helios 2 \\

10 &         &  78,79  &  22 -- 08  &  0.57  & Helios 1 \\  

11 &         &  159,160  &  18 -- 13  &  0.86  & Helios 1 \\ 

12 &         &  240,241  &  14 -- 10  &  0.84  & Helios 1 \\   

13 &         &  268,269  &  14 -- 12  &  0.57  & Helios 1 \\ 

14 &         &  335  &  14 -- 00  &  0.75  & Helios 1 \\

15 &  1978   &  3,4  &  15 -- 17  &  0.95  & Helios 1 \\ 

16 &         &  4,5  &  08 -- 10  &  0.94  & Helios 2 \\ 

17 &         &  6  &  01 -- 13  &  0.95  & Helios 2 \\

18 &         &  17  &  01 -- 23  &  0.98  & Helios 2 \\  

19 &         &  29,30  &  12 -- 01  &  0.98  & Helios 2 \\ 

20 &         &  37,38  &  16 -- 16  &  0.98  & Helios 2 \\ 

21 &         &  46,47  &  14 -- 20  &  0.95  & Helios 1 \\ 

22 &         &  47,48  &  03 -- 09  &  0.95  & Helios 2 \\ 

23 &         &  61,62  &  01 -- 01  &  0.87  & Helios 1 \\

24 &         &  92  &  02 -- 07  &  0.61  & Helios 2 \\  

25 &         &  114  &  12 -- 19  &  0.32  & Helios 2 \\

26 &         &  189,190  &  22 -- 22  &  0.94  & Helios 1 \\ 

27 &         &  292,293  &  01 -- 14  &  0.47  & Helios 1 \\ 

28 &         &  358,359  &  15 -- 15  &  0.85  & Helios 2 \\  

29 &         &  363,364  &  09 -- 14  &  0.85  & Helios 1 \\ 

30 &  1979   &  58,59  &  15 -- 15  &  0.96  & Helios 1 \\ 

31 &         &  62  &  09 -- 17  &  0.94  & Helios 1 \\

32 &         &  93  &  02 -- 18  &  0.68  & Helios 2 \\

33 &         &  129  &  06 -- 12  &  0.30  & Helios 2 \\

34 &         &  148,149  &  23 -- 07  &  0.43  & Helios 1 \\

35 &         &  305  &  03 -- 19  &  0.50  & Helios 1 \\ 

36 &  1980   &  82,83  &  17 -- 12  &  0.92  & Helios 1 \\     

37 &         &  90  &  01 -- 15  &  0.88  & Helios 1 \\ 

38 &         &  162,163  &  17 -- 01  &  0.41  & Helios 1 \\

39 &         &  172  &  02 -- 20  &  0.53  & Helios 1 \\ 

40 &         &  175  &  05 -- 17  &  0.57  & Helios 1 \\ 

41 &         &  231  &  00 -- 18  &  0.97  & Helios 1 \\ 

42 &         &  117,118  &  09 -- 03  &  0.79  & Helios 1 \\

43 &         &  131,132  &  15 -- 03  &  0.66  & Helios 1 \\   

44 &         &  146,147  &  03 -- 07  &  0.48  & Helios 1 \\ 

45 &         &  170  &  03 -- 09  &  0.34  & Helios 1 \\ 

 \hline
 
 \end{tabular}
 
 \end{center}
 
 \end{table}
 


\bibliography{sola_draft_1}

\end{article} 

\end{document}